\documentclass[prd,twocolumn,showpacs,superscriptaddress,nofootinbib,floatfix,showkeys,10pt]{revtex4-1}
\usepackage{graphicx}
\usepackage{amsmath}
\usepackage{bm}
\usepackage{yhmath}
\usepackage{mathtools}
\usepackage{wasysym}
\usepackage[colorlinks,citecolor=blue,urlcolor=blue,linkcolor=blue]{hyperref}
\usepackage{subfigure}
\usepackage{color}
\usepackage{cases}
\usepackage{subfigure}
\usepackage{times}
\usepackage{dcolumn,booktabs,bm}
\usepackage{slashed}
\usepackage{amsfonts,amssymb,stmaryrd,latexsym,amsmath}
\usepackage{textcomp}
\usepackage{multirow}
\usepackage{cancel}
\usepackage{array}
\usepackage{orcidlink}

\def\SU3{{\text{SU(3)}_{\rm F}}}

\def \pcs4338{{P_{\psi s}^\Lambda(4338)^0}}

\allowdisplaybreaks

\begin{document}

\title{The double thresholds distort the lineshapes of the $P_{\psi s}^\Lambda(4338)^0$ resonance}

\author{Lu Meng\,\orcidlink{0000-0001-9791-7138}}
\affiliation{Ruhr-Universit\"at Bochum, Fakult\"at f\"ur Physik und
Astronomie, Institut f\"ur Theoretische Physik II, D-44780 Bochum,
Germany }

\author{Bo Wang\,\orcidlink{0000-0003-0985-2958}}\email{wangbo@hbu.edu.cn}
\affiliation{School of Physical Science and Technology, Hebei
    University, Baoding 071002, China}
\affiliation{Key Laboratory of High-precision Computation and
    Application of Quantum Field Theory of Hebei Province, Baoding
    071002, China}

\author{Shi-Lin Zhu\,\orcidlink{0000-0002-4055-6906}}\email{zhusl@pku.edu.cn}
\affiliation{School of Physics and Center of High Energy Physics,
Peking University, Beijing 100871, China}


\begin{abstract}
Very recently, the LHCb Collaboration reported the first observation of the hidden charm pentaquark with strangeness, $P_{\psi s}^\Lambda(4338)^0$. Considering this state is very close to the $\Xi_c^0\bar{D}^0$ and $\Xi_c^+D^-$ thresholds, we explore the possible bias of the Breit-Wigner parameterization, with emphasis on the effect of its coupling to the double thresholds $\Xi_c^0\bar{D}^0$ and $\Xi_c^+D^-$. We first use a qualitative picture based on the ``uniformization" of the Riemann surface of the two-channel system to understand the positions of the enhancement. Then we use the Lippmann-Schwinger equation (LSE) formalism (equivalent to the $K$-matrix parameterization) with two models, the zero-range model and Flatt\'e model to investigate the  $J/\psi \Lambda$ lineshapes. Our results show that the nominal peak of the $P_{\psi s}^\Lambda(4338)^0$ could arise either from the pole well above the $\Xi_c^+ D^-$ threshold on the $(-,+)$ sheet or from the pole well below the $\Xi_c^0 \bar{D}^0$ threshold on the $(-,-)$ sheet in the two-channel system. Using the Breit-Wigner distribution to depict the above two lineshapes could be misleading. We also find a novel type of lineshapes with the enhancement constrained by the threshold difference. We urge the LHCb collaboration to perform the refined experimental analysis considering the unitarity and analyticity, e.g. using the K-matrix parameterization. As a by-product, we obtain that the ratio of the isospin violating decay ${\Gamma_{P_{\psi s}^{\Lambda}\to J/\psi\Sigma}}/{\Gamma_{P_{\psi s}^{\Lambda}\to J/\psi\Lambda}}$ could be up to $10\%$.

\end{abstract}

\maketitle

\thispagestyle{empty}

\section{Introduction}

\begin{figure}[htbp]
    \centering  \includegraphics[width=0.3\textwidth]{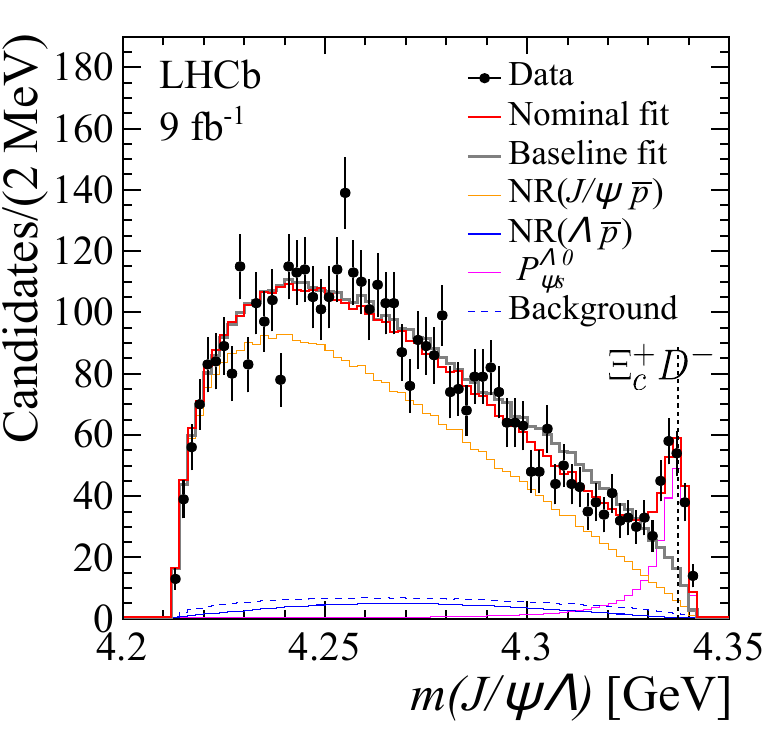}
    \caption{The experimental fitting with the relativistic Breit-Wigner lineshape of the $\pcs4338$~\cite{LHCb:2022jad}.}\label{fig:exlineshape}
\end{figure}

Very recently, the LHCb Collaboration announced the first observation of the hidden-charm pentaquark state  with strangeness~\cite{LHCb:2022jad}. The signal was observed  in the $J/\psi \Lambda$ invariant mass spectrum of the decay $B^-\to J/\psi \Lambda \bar{p}$. The state is composed of at least five quarks $(c\bar{c}uds)$. Within the new naming convention recommended by the LHCb Collaboration~\cite{Gershon:2022xnn} (the convention will be adopted here and following), the state is labeled as $\pcs4338$. Within a relativistic Breit-Wigner (BW) lineshape fitting as shown in Fig.~\ref{fig:exlineshape}, the mass and width of the resonance were extracted,
\begin{eqnarray}
    m&=&4338.3\pm 0.7\pm 0.4\text{ MeV},\nonumber\\
    \Gamma &=&7.0\pm 1.2 \pm 1.3 \text{ MeV}.
\end{eqnarray}
One can see that the resonance is very close to the $\Xi_{c} \bar{D}$ thresholds ($\sim4336$ MeV). Meanwhile, the amplitude analysis prefers the ${1\over 2}^-$ spin-parity quantum numbers and excludes the possibility of ${1\over 2}^+$ at 90\% confidence level.

The observation of $\pcs4338$ is the
follow-up story of the $P_{\psi}^N$ states~\cite{LHCb:2015yax,LHCb:2019kea}, see~\cite{Chen:2016qju,Liu:2019zoy,Guo:2017jvc,Lebed:2016hpi,Brambilla:2019esw,Chen:2022asf,Meng:2022ozq} for recent reviews of the exotic states. The evidence of the pentaquark with strangeness, $P_{\psi s}^\Lambda (4459)^0$ was reported by LHCb~\cite{LHCb:2020jpq}, but the significance is less than $5~\sigma$. After the observation of the $\pcs4338$, Karliner and Rosner pointed out that the proximity to the threshold, its $J^P={1\over 2}^-$ quantum numbers and the narrow width strongly suggest its $\Xi_c \bar{D}$ molecule nature~\cite{Karliner:2022erb}. The partners of the $\pcs4338$ were also investigated in Refs.~\cite{Wang:2022mxy,Yan:2022wuz,Wang:2022neq}. It
is worthwhile to emphasize that the $\pcs4338$ had been predicted as the $\Xi_c \bar{D}$ molecule before the experimental report. In 2019, we investigated the spectrum of the strange hidden charm molecular pentaquarks systematically with chiral effective field theory in Ref.~\cite{Wang:2019nvm} (see~\cite{Meng:2022ozq} for a recent review). The $\Xi_c \bar{D}$ bound state with quantum number $I(J^P)=0({1\over 2}^-)$ was predicted. In the same work, the $\Xi_c\bar{D}^*$ bound state was also obtained, which coincides to the experimental $P_{\psi s}^\Lambda (4459)^0$ state. Later, a unified description of the loosely bound molecular systems composed of the heavy flavor hadrons $(\bar{D},\bar{D}^*)$, $(\Lambda_c, \Sigma_c, \Sigma_c^*)$, and
$(\Xi_c, \Xi_c^\prime,\Xi_c^*)$ was presented
in Ref.~\cite{Chen:2021cfl}, where the $\Xi_c \bar{D}$ molecular state was predicted around 4328 MeV. Either the $\Xi_c \bar{D}$ bound state or virtual state depending on the cutoff parameter was predicted in the vector-meson-exchange model in Ref.~\cite{Dong:2021juy}. Before the experimental results of LHCb, the hidden charm pentaquarks with strangeness, were also investigated in Refs.~\cite{Wu:2010jy,Chen:2016ryt,Feijoo:2015kts,Lu:2016roh,Shen:2019evi,Xiao:2019gjd}. The evidences of $P_{\psi s}^\Lambda(4459)^0$ incited a new round of discussion~\cite{Chen:2020uif,Gao:2021hmv,Ozdem:2021ugy,Du:2021bgb,Peng:2020hql,Li:2021ryu,Hu:2021nvs}.

The peak of the resonance in Fig.~\ref{fig:exlineshape} is in the vicinity of the $\Xi_c\bar{D}$ thresholds. Specifically, the mass extracted from the relativistic BW parameterization is about 0.93 MeV above the $\Xi_c^+ D^-$ threshold as shown in Fig.~\ref{fig:ppole}. Considering the uncertainty of the mass, it is hard to judge whether this signal is a below-threshold or above-threshold state.  For a long time, it has been realized that the lineshape of the resonance would be distorted from the conventional BW distribution if it appears near the threshold and  strongly couples to the threshold at the same time
~\cite{Flatte:1976xu,Hanhart:2007yq,Hanhart:2015cua}. Therefore, the present BW mass and width could have large discrepancy to the pole position. What is more, the effect of the $\Xi_c^0 \bar{D}^0$ threshold could also be important. It is about 3 MeV below the BW mass, which is comparable to the half width of the resonance.  In principle, the enhancement in the lineshape could arise from the pure kinetic effect, such as the threshold effect and triangle singularity (see Ref.~\cite{Guo:2019twa} for a comprehensive review), rather than from the pole of the $T$-matrix. However, it rarely happens. The more common case is that the resonance lineshape is distorted by the threshold effect, or equivalently, the threshold effect is amplified by the nearby pole. Therefore, in the work, we focus on the distorting effect of the double thresholds on the resonance lineshape. In contrast to the literature concerning one threshold, the effect of the double thresholds will  be emphasized. We will first discuss the uniformization~\cite{newton1982scattering,Yamada:2020rpd,Yamada:2021cjo,Yamada:2021azg,Yamada:2022xam} technique to unfold the Riemann sheets and present a qualitative picture of the double threshold effect. Then we will use two specific models to show the lineshapes explicitly. 

In Sec.~\ref{sec:uniform}, we analyze the topological structure of the two-channel $T$-matrix and introduce the uniformization of the two-channel Riemann surface. In Sec.~\ref{sec:models}, we introduce the formalism to investigate the lineshape of the $J/\psi \Lambda$ in the $B^-\to J/\psi \Lambda  \bar{p}$ decay with two models to depict the $J/\psi \Lambda$-$\Xi_c^0\bar{D}^0$-$\Xi_c^+D^-$ rescattering effect. In Sec.~\ref{sec:lineshape}, we choose twelve different pole masses on different Riemann sheets to show the possible lineshapes of the resonances. In Sec.~\ref{sec:sum}, we give a brief summary. In Appendix~\ref{app:two_model}, we will clarify the relations between our two models in this work and those in literature. In Appendix~\ref{app:iso_v}, we evaluate the isospin violating decay $\pcs4338 \to J/\psi \Sigma$.

\section{Uniformization of the two-channel Riemann surface}~\label{sec:uniform}

In this work, we will focus on three channels,
\begin{equation}
|1\rangle=|J/\psi\Lambda\rangle,\quad|2\rangle=|\Xi_{c}^{0}\bar{D}^{0}\rangle,\quad|3\rangle=|\text{\ensuremath{\Xi_{c}^{+}D^{-}}\ensuremath{\ensuremath{\rangle}}}. \label{eq:channels}
\end{equation}
The threshold of the corresponding channel $|i\rangle $ is labelled as $m_{T_i}$. The resonance is in the vicinity of the $m_{T_2}$  and $m_{T_3}$, which is our energy region of interest. However, the $m_{T_1}$ is far below the energy region, $m_{T_{2,3}}-m_{T_{1}}\sim 125$ MeV, which is also about the $Q$-value of $\pcs4338\to J/\psi \Lambda$ .
Very similar to the hidden-charm decays of
the $P_{\psi}^N$ states, such a large $Q$-value only induced a small width about 7 MeV, which implies that the coupling between $\pcs4338$ and $J/\psi \Lambda$ is very weak and supports the $\Xi_c \bar{D}$ molecular interpretation of the $\pcs4338$. Arranging two well-separated $c$ and $\bar{c}$ in the hadronic molecule into a single meson ($J/\psi$) is suppressed naturally. Similar mechanisms are also responsible for the dominant decay patterns of the charmonium-like states~\cite{Meng:2021rdg,Meng:2021kmi,Meng:2020ihj}. Based on the above analyses, it is rational to assume that the $\pcs4338$ is generated from the $\Xi_c\bar{D}$ interaction. The inclusion of the $J/\psi \Lambda $ channel will not affect its existence but slightly correct its pole position. In the following, we will first focus on the two-channel problem ($|\Xi_c^+D^-\rangle$ and $|\Xi_c^0\bar{D}^0\rangle$ channels).

\begin{figure}[htp]
    \centering  \includegraphics[width=0.5\textwidth]{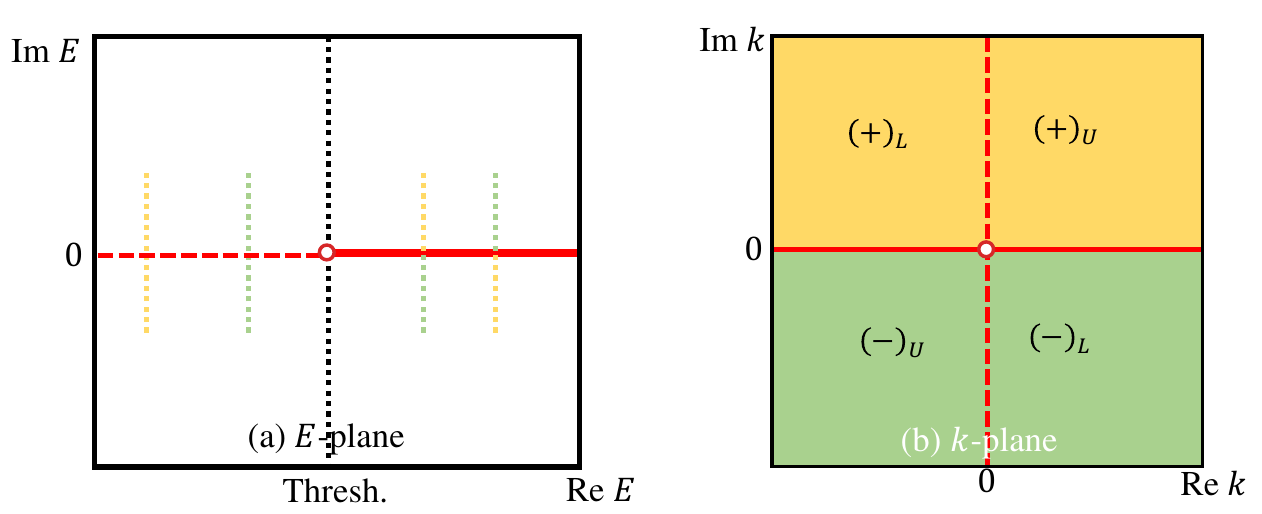}
    \caption{The topological structure of Riemann surface and its uniformization for the single-channel system. The different sheets are shown in different colors, with label of signs of the imaginary parts of momenta. The subscript ``U" and ``L" represent upper- and lower-half sheets respectively. The red solid line, red dashed line and red open markers represent the $k^2 >0$, $k^2<0$ and $k^2=0$, respectively. The subfigure (a) illustrates connection relations of different sheets by the colored lines cross real axis in different regions.}\label{fig:unif1}
\end{figure}

\begin{figure}[htp]
    \centering  \includegraphics[width=0.5\textwidth]{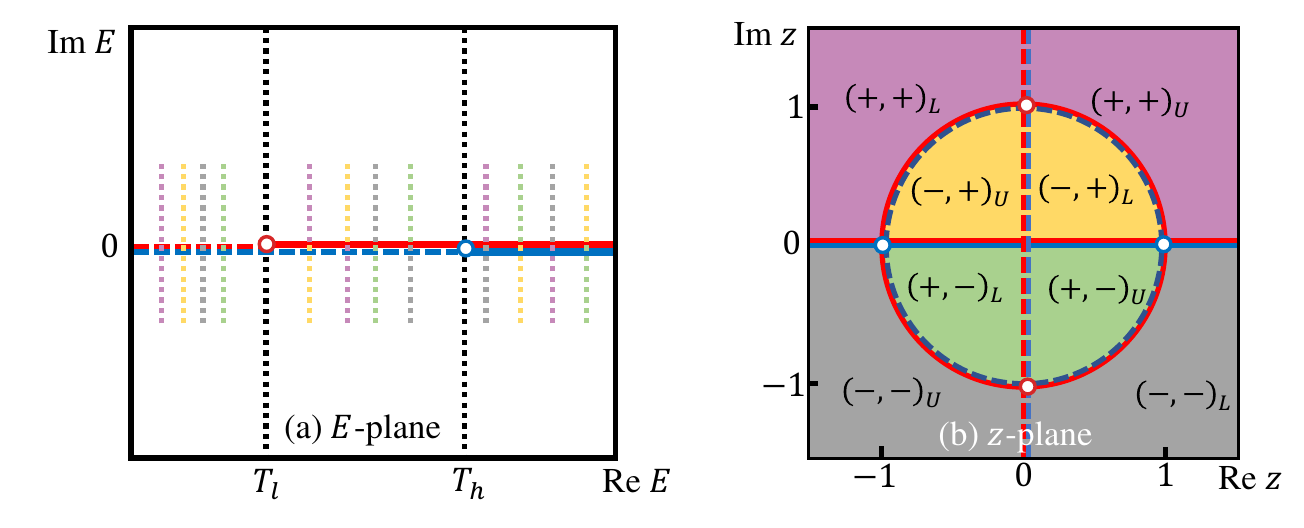}
    \caption{The topological structure of Riemann surface and its uniformization for the double-channel system. The different sheets are shown in different colors, with label of signs of the imaginary parts of momenta. The subscript ``U" and ``L" represent upper- and lower-half sheets respectively. The solid lines, dashed lines and open markers represent the $k_i^2 >0$, $k_i^2<0$ and $k_i^2=0$ with red (blue) for the lower (higher) channel, respectively. The subfigure (a) illustrates connection relations of different sheets by the colored lines cross real axis in different regions.}\label{fig:unif2}
\end{figure}

\begin{figure}[htp]
    \centering  \includegraphics[width=0.45\textwidth]{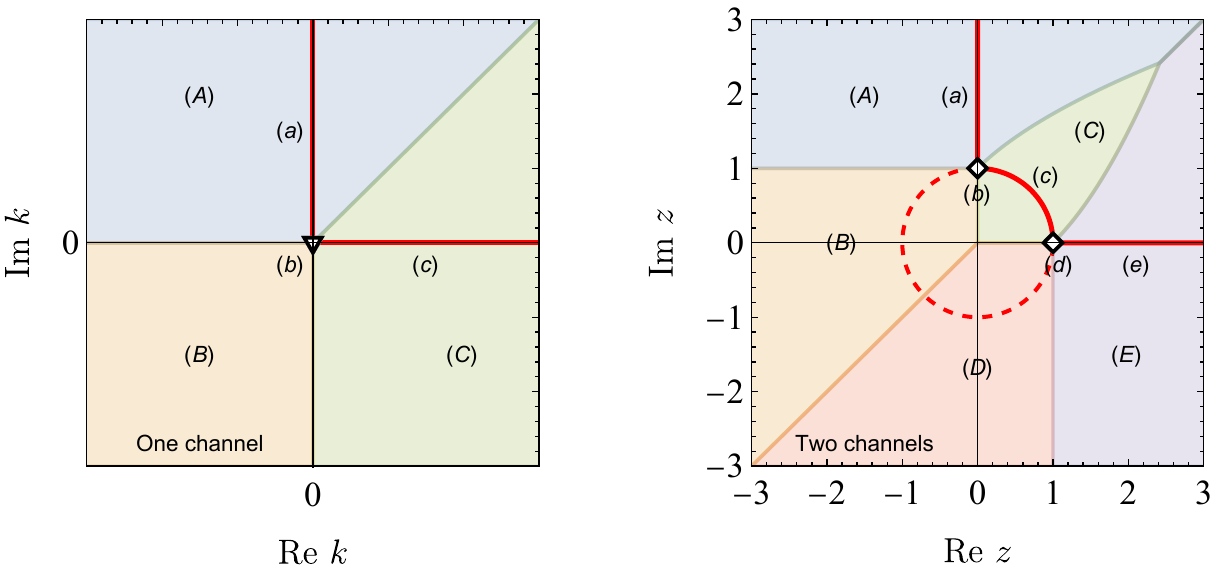}
    \caption{The corresponding relations between the pole position and $|T|^2$ peak position in the physical region under the assumption in Eqs.~\eqref{eq:pole_k} and \eqref{eq:pole_z}. For each subfigure, the regions marked with different colors [or $(A),(B),\dots$] imply that the distance from the poles in these regions to the physical regions [$(a),(b),...$] is the shortest. Consequently, the pole at the regions $(A),(B),...$ will give rise to a peak at the physical regions $(a),(b),...$, respectively. The left and right subfigures are for the one-channel system  and two-channel system, respectively.}\label{fig:region}
\end{figure}

In general, the elements of the multi-channel $S$-matrix are functions of momentum $k_i$ of each channel. Because of the square-root type function relating $k_i$ to energy $E$, the elements of the $S$-matrix become the multi-value functions in the complex $E$-plane. With the opening of each channel, an extra branch of the $S$-matrix comes up. The branch cut is related to the unitarity of the $S$-matrix and the starting point (branch point) of the cut is just the threshold of the new opening channel. In Fig.~\ref{fig:unif1} (a) and Fig.~\ref{fig:unif2} (a), we illustrate the topological structures of the Riemann surfaces of the single-channel and two-channel systems, respectively. We use the signs of the imaginary parts of the momenta in the threshold ascending order to label different  Riemann sheets~\cite{Eden:1964zz,Frazer:1964zz,Badalian:1981xj}. For example, there are four Riemann sheets for two-channel system,  $(+,+)$, $(-,+)$, $(-,-)$ and $(+,-)$, where the first and the second signs are for the lower and higher channels, respectively. We also introduce the subscript ``U" and ``L" to label the upper-half and lower-half planes respectively. The physical region appears on the real axis of the physical sheet with all positive signs.

If the resonance pole is near the physical region but away from the branch points, the pole surrounding could be regarded as flat and one can use a simple pole of $E$ to parameterize the resonance,
\begin{equation}
    T\propto {1 \over E-E_0}\propto  {1 \over E-M+i\Gamma/2},
\end{equation}
 where $E_0$ is the pole position with $M$ and $-\Gamma/2$ as the real and imaginary parts. This is the nonrelativistic BW parameterization. However, when the resonance pole appears near the thresholds, the branching behavior of $T(E)$ will make the simple parameterization unreasonable. Alternatively, we could try to find a variable $z$ to set a mapping from $E$ to $z$ and make the $T$-matrix a single value function of $z$ in the locally flat surface. This process is called uniformization.  It has been shown that the single-channel and two-channel $S$-matrix can be mapped into a single plane~\cite{newton1982scattering,Yamada:2020rpd,Yamada:2021cjo,Yamada:2021azg}.

For the single-channel system, the most convenient uniformization is to go to the momentum plane. In Fig.~\ref{fig:unif1} (b), we present the $k$-plane of the $S$-matrix. For the two-channel system, we first introduce two momentum-like variables $q_l$ and $q_h$,
 \begin{equation}
     q_{l}^{2}=(E^{2}-m_{T_{l}}^{2}),\quad q_{h}^{2}=(E^{2}-m_{T_{h}}^{2}),
 \end{equation}
where $m_{T_l}$ and $m_{T_h}$ are the lower and higher thresholds, respectively. For our problem, there are $m_{T_l}=m_{T_2}$ and $m_{T_h}=m_{T_3}$. Apparently,  the sign of the imaginary part of $q_i~(i=l,h)$ should be related to different sheets of $S(E)$. We can introduce the real positive $\Delta$ as,
 \begin{equation}
     q_{l}^{2}-q_{h}^{2}=m_{T_{h}}^{2}-m_{T_{l}}^{2}\equiv\Delta^{2}.
 \end{equation}
 We set up the mapping $E$ to $z$ from the following relations,
 \begin{equation}
     q_{l}+q_{h}=\Delta z,\quad q_{l}-q_{h}=\frac{\Delta}{z}.\label{eq:uniformap}
 \end{equation}
In Fig.~\ref{fig:unif2} (b), we present the $z$-plane of  the $S$-matrix, where the regions corresponding to four sheets and two cuts are shown in different colors. Apparently, the element of the $S$-matrix will be a single value function of the $z$. The two branch points in the $E$-plane is unfolded into four points, which are shown as open markers in  Fig.~\ref{fig:unif2} (b). The origin point corresponds to the infinities  of $(-,+)$ and $(+,-)$ sheets. For the non-relativistic system, one can introduce $q_{l,h}$ as follows,
\begin{equation}
    q_{i}^{2}\equiv(m_{T_{h}}+m_{T_{l}})(E-m_{T_{i}})=(m_{T_{h}}+m_{T_{l}})\frac{k_{i}^{2}}{2\mu_{i}},
\end{equation}
where $\mu_{i}$ is the reduced mass for the corresponding channel. The remaining derivations are the same as those of the relativistic case.

In Fig.~\ref{fig:region}, we show the physical regions of the single-channel and double-channel systems in the uniformized plane with the solid red lines. For the single-channel system, the positive imaginary axis $(a)$ corresponding to the $E<0$ region. The positive real axis $(c)$ corresponding to the cut in the physical regions. The two lines form a right angle with vertex of the branch point $(b)$. Assuming a smooth function in the $k$-plane, its value along the physical region $(a)\to (b) \to (c) $ will become unsmooth at point $(b)$. In this picture, one can easily understand the appearance of the ``cusp" effect in the threshold. For the two-channel system, the physical regions are $(a)$ where neither channel opens, $(b)$ the lower threshold, $(c)$ where the only the lower channel opens, $(d)$ the higher threshold and $(e)$ where both thresholds open. The lines or arc form two right angles at two thresholds. Thus, one can expect the unsmoothness appears at two thresholds for the amplitudes. It should be noticed that one could choose other uniformizations e.g. by introducing another mapping with the arbitrary analytical function $g(k)$ or $g(z)$. However, the transformation introduced by the analytical function will be conformal, which locally prevents the angle. Thus, the above discussion about the threshold effects based on the right angles will not change.

Now, we expect the $T$-matrix will be the analytical function of $k$ or $z$ after uniformization except possible poles. Naively, we could introduce the simplest pole parametrizations like the Breit-Wigner function but in $k$- or $z$-plane.
\begin{eqnarray}
   \text{single-channel: } T &\propto& {1 \over (k-k_0)} \propto {1 \over (k-k_r-ik_i)},~\label{eq:pole_k}\\
    \text{double-channels: } T &\propto & {1 \over (z-z_0)} \propto {1 \over (z-z_r-iz_i)}.~\label{eq:pole_z}
\end{eqnarray}
where $k_0=k_r+ik_i$ and $z_0=z_r+iz_i$ are poles. Apparently, the $|T|^2$ is inversely proportional to the square of the geometric distance between $z$($k$) and $z_0$ ($k$) in the uniformized plane. Here we only keep the contribution of the single pole. With this rough picture, we know the physical $|T|^2$ will achieve its maximum at the point closest to the pole. In Fig.~\ref{fig:region}, we divide the whole planes into several regions  $(A),(B),...$ according to their closest physical regions $(a),(b),...$. For example, the pole appears in region $(A)$ will give a peak at physical region $(a)$. Comparing Fig.~\ref{fig:region} with Fig.~\ref{fig:unif1} (b) and Fig.~\ref{fig:unif2} (b), one can get a rough impression where the peak will appear for the pole at different sheets.

For the single channel system, the pole on the $(-)_L$ sheet, the lower-half $E$-plane of the second sheet, will give a peak above the threshold. For every pole on the $(-)_L$ sheet, there is a conjugate pole at $(-)_U$ sheet (we will discuss it in details later). Such a pole could enhance the threshold effect. The virtual state pole on the negative imaginary axis of $k$-plane tends to contribute to a peak on the threshold.

For the two-channel system, the pole on the $(+,-)$ or $(-,+)$ sheets only give a peak between two thresholds (including two thresholds). The large region (B) and (D)  will give a peak in thresholds, which include the whole $(+,-)$ sheet, upper-half $(+,-)$ region, part of $(+,+)$ sheet and part of $(-,-)$ sheet. Meanwhile, one can see the pole on the $(-,-)$ sheet only gives a peak above the lower threshold. The pole on the $(+,+)$ sheet could give a  potential peak in all $(a)-(e)$ regions.

Apart from the above topological property, the analyticity constrains the $S$-matrix from the Schwartz reflection principle~\cite{Taylor:1972pty}. Apparently, the elements of the $S$-matrix in the $(a)$ region in Fig.~\ref{fig:region} are real and analytical except the possible bound state poles. With the Schwartz reflection principle, the elements of the $S$-matrix satisfy
\begin{equation}
 S(z)=S^*(-z^*),\label{eq:reflection}
\end{equation}
where $z$ is the variable after uniformization and becomes $k$ for the single-channel systems. Considering the $T\sim (1-S)/(ik_1)$, the matrix elements of $T(z)$ satisfy the same reflection rule and the poles will appear in pairs as
\begin{equation}
    T(z)=\frac{c_{0}}{z-z_{0}}-\frac{c_{0}^{*}}{z+z_{0}^{*}},~\label{eq:twopole}
\end{equation}
where $c_0$ and $-c_0^*$ are the residues of the two poles. In the $E$-plane, the pole will become symmetric with respect to the real axis for each sheet. If we assume $c_0$ is real $c_0=c_0^*$, we can get the possible lineshapes of the $|T(z)|^2$.

\begin{figure*}[htp]
    \centering  \includegraphics[width=0.9\textwidth]{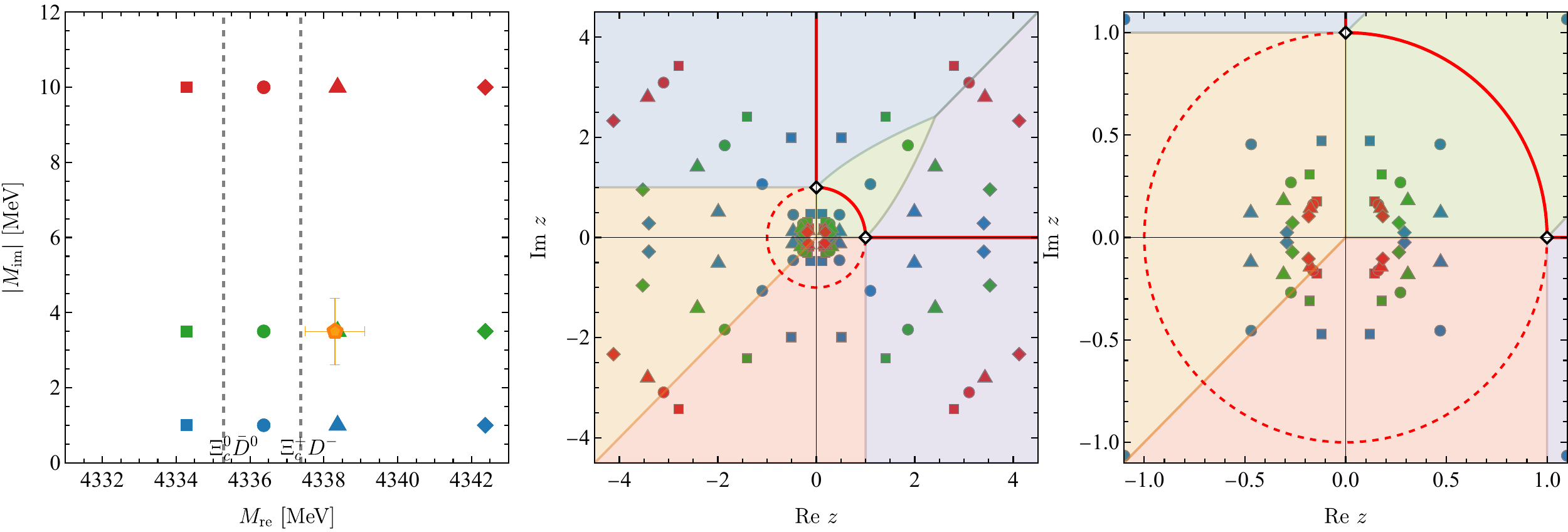}
    \caption{The ``synthetic''  pole positions, the $\Xi_c\bar{D}$ thresholds and experimental mass and width. In the left subfigure, the gray dashed lines represent two $\Xi_c\bar{D}$ thresholds. The orange pentagon with error bars stands for the experimental measurement of the mass and width of $\pcs4338$ within relativistic BW parameterization. The other markers are the ``synthetic''  poles used to investigate the lineshapes. The middle and right subfigures are the distribution of these poles on the $z$-plane.}\label{fig:ppole}
\end{figure*}

\begin{figure*}[htbp]
    \centering  \includegraphics[width=0.9\textwidth]{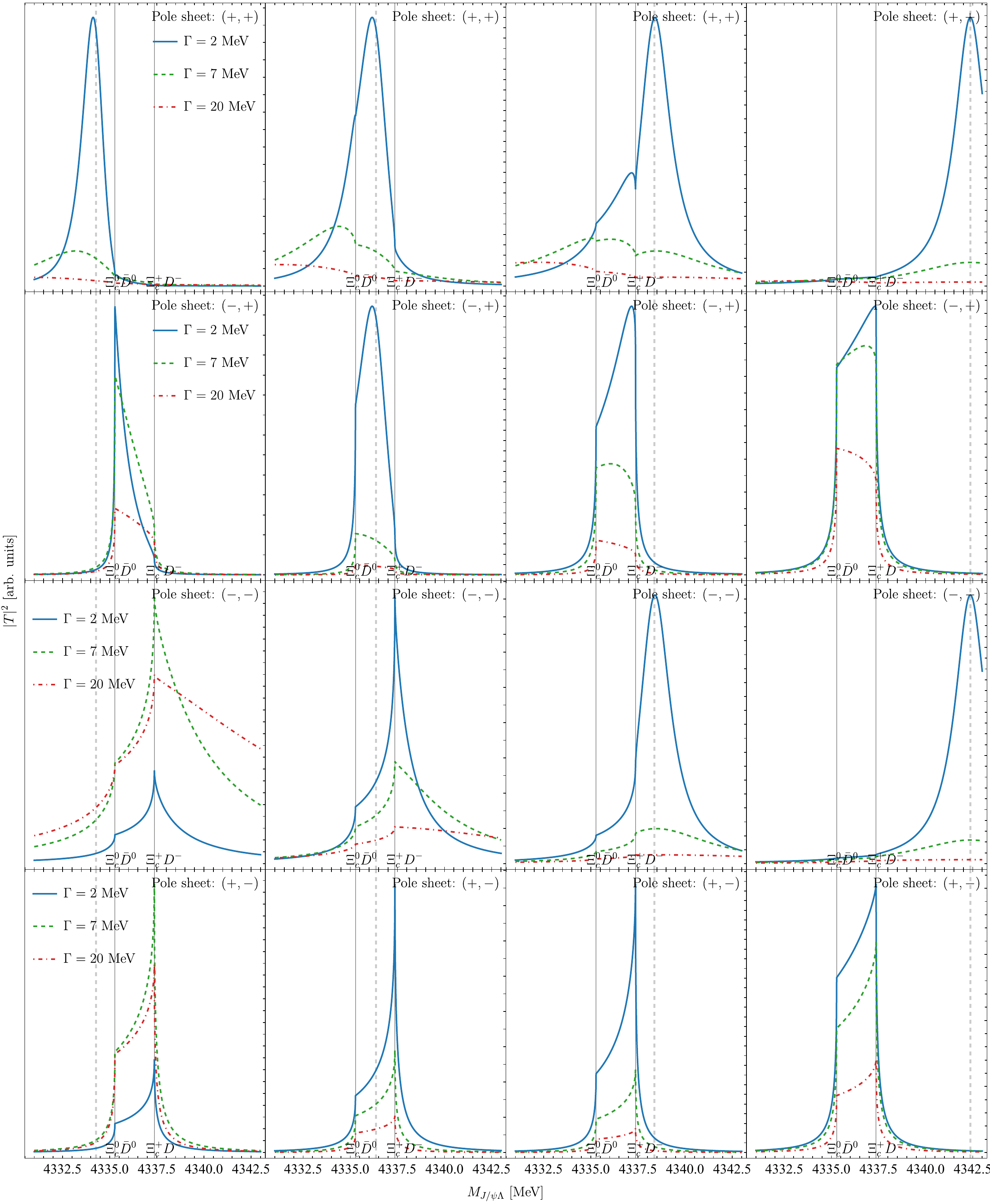}
    \caption{The lineshape of $|T|^2$ using the parameterization in Eq.~\eqref{eq:twopole} with $c_0=c_0^*$.}\label{fig:polecontri}
\end{figure*}

We take twelve different pole masses $M_{i,j}=M_\text{re}^{i}\pm\mathrm{i}M_\text{im}^{j}$ with $i=0,1,2,3$ and $j=1,2,3$ to investigate the corresponding lineshapes, with
\begin{eqnarray}
    M_\text{re}&\in&\{(m_{T_{2}}-1\text{MeV}),\;(m_{T_{3}}-1\text{MeV}),\nonumber \\
    &&~~~~(m_{T_{3}}+1\text{MeV}),\;(m_{T_{3}}+5\text{MeV})\}, \nonumber\\
    |2M_\text{im}|&=&\Gamma\in\{2,\;7,\;20\}\text{MeV}. \label{eq:polemass}
\end{eqnarray}
These ``synthetic" poles are presented in Fig.~\ref{fig:ppole}. One can see that these poles are below the two thresholds, between the two thresholds, slightly above the higher threshold, and significantly above the higher thresholds. The imaginary values include the experimental one (treating BW mass as the pole mass),  a much smaller one and much larger one. In the calculation, we will consider the twelve poles on different Riemann sheets. In principle, we have taken all the qualitatively different cases into consideration. We plot the lineshape of $|T|^2$ in Fig.~\ref{fig:polecontri} using parametrization in Eq.~\eqref{eq:twopole} with $c_0=c_0^*$.

One can see that the pole appearing in $(+,-)$ and $(-,+)$ sheets will give an enhancement to the region between two thresholds, which agrees with  Fig.~\ref{fig:region}. The peaks move from the lower threshold to the higher threshold with the increasing mass for the $(-,+)$ sheet poles. However, the peaks only appear on the second threshold for the $(+,-)$ sheet poles. For the poles on the $(-,-)$ sheet, the peaks appear in the regions above the higher thresholds. When the pole mass is below the higher threshold, the peak appears on the second threshold. When the pole mass is increasing above the higher threshold, the peak tends to move with the pole mass. Only for the poles on the $(+,+)$ sheet with a small imaginary part, the peak will move with the pole mass all the time. Therefore, in most cases, the BW mass could not reflect the real pole position. We will see the results from two dynamical models are qualitatively consistent with the analysis with the simple parameterization in the uniformization scheme.

\color{black}

\section{Dynamical models}~\label{sec:models}
In order to verify the qualitative analysis in Sec.~\ref{sec:uniform}, we adopt two models to calculate the lineshapes explicitly. In the model-I, we introduce the contact interactions for three channels in Eq.~\eqref{eq:channels} with the isospin symmetry,
\begin{eqnarray}
V_{\text{I}}=\frac{1}{2}\left[\begin{array}{ccc}
    0 & -\tilde{c} & \tilde{c}\\
    -\tilde{c} & c_{1}+c_{0} & c_{1}-c_{0}\\
    \tilde{c} & c_{1}-c_{0} & c_{1}+c_{0}
\end{array}\right],~\label{eq:model-I}
\end{eqnarray}
where $c_1$, $c_0$ and $\tilde{c}$ are defined as follows,
\begin{eqnarray}
    \langle\Xi_{c}D,I=i|\hat{V}|\Xi_{c}D,I=j\rangle\equiv c_{i}\delta_{ij},\\
    \langle\Xi_{c}D,I=0|\hat{V}|J\psi\Lambda,I=0\rangle\equiv \tilde{c}.
\end{eqnarray}
We omit the interaction between $J/\psi$ and $ \Lambda$. The above interaction is a natural extension of our previous works~\cite{Meng:2021kmi,Meng:2021rdg,Meng:2021jnw,Meng:2020cbk,Meng:2020ihj}.

In the model-II, the $\Xi_c \bar{D}$ interactions are introduced through a bare isospin singlet resonance which couples to these $\Xi_c \bar{D}$ channels,
\begin{eqnarray}
    V_{\text{II}}=\frac{1}{2}\left[\begin{array}{ccc}
        0 & -\tilde{c} & \tilde{c}\\
        -\tilde{c} & \frac{g^{2}}{E^2-m_{0}^{2}} & -\frac{g^{2}}{E^2-m_{0}^{2}}\\
        \tilde{c} & -\frac{g^{2}}{E^2-m_{0}^{2}} & \frac{g^{2}}{E^2-m_{0}^{2}}
    \end{array}\right],~\label{eq:model-II}
\end{eqnarray}
where $m_0$ is the bare mass of the resonance and $g$ is the coupling constant. The coupling of $J/\psi \Lambda$ and $\Xi_c\bar{D}$ is the same as that in the model-I.

With the interactions, the coupled-channel $T$-matrix can be obtained by solving the LSEs,
\begin{equation}
    T=V+VGT,\quad G=\text{diag}\{G_{1},G_{2},G_{3}\}.~\label{eq:LSE}
\end{equation}
The $G_i$ is
\begin{eqnarray}
    G_{i}(E)&=i\int\frac{d^{4}l}{(2\pi)^{4}}\frac{1}{l^{2}-m_{i1}^{2}+i\epsilon}\frac{1}{(P-l)^{2}-m_{i2}^{2}+i\epsilon}\\
    &=\int_{0}^{\Lambda}\frac{l^{2}dl}{(2\pi)^{2}}\frac{\omega_{i1}+\omega_{i2}}{\omega_{i1}\omega_{i2}[E^{2}-(\omega_{i1}+\omega_{i2})^{2}+i\epsilon]},
\end{eqnarray}
with $\omega_{ia}=(\bm{l}^{2}+m_{ia}^{2})^{1/2}$. We use the $m_{ia}$ to denote the mass of the $a$-th particle of the $i$-th channel. The total momentum of the two particles reads $P=(E,\bm{0})$ at the center of the mass frame. The analytical results of the above integral can be found in Ref.~\cite{Oller:1998hw}. In order to continue the $T$-matrix to the unphysical sheets, one can use the following replacement to the channels with a negative imaginary part of the momentum,
\begin{equation}
G_{i}\to G_{i}+i\frac{k_{i}}{4\pi E},
\end{equation}
where the $k_i$ is defined as
\begin{eqnarray}
k_{i}=\frac{\sqrt{[E^{2}-(m_{i1}+m_{i2})^{2}][E^{2}-(m_{i1}-m_{i2})^{2}]}}{2E}. \label{eq:k}
\end{eqnarray}
The physical meaning is the momentum of the final states $m_{i1}$ or $m_{i2}$ in the two body decay of a mother particle with mass $E$. In the calculation, we keep the slight mass differences of the charged and neutral $\Xi_c~(\bar{D})$ to embed the possible isospin violation effect. In our calculation, the reflection principle in Eq.~\eqref{eq:reflection} is satisfied.

In order to show the relations of above models with the other similar ones in literature, we give the reduced non-relativistic interaction of the $\Xi_c\bar{D}$ two-channel systems in Appendix~\ref{app:two_model}. One can see the clear corresponding relations between the LSE formalism and the $K$-matrix parameterization~\cite{ParticleDataGroup:2020ssz}.  Meanwhile, one can see that the model-I is the zero-range model in Refs.~\cite{Braaten:2007nq,Artoisenet:2010va}. The model-II is the Flatt\'e parameterization in Refs.~\cite{Flatte:1976xu,Hanhart:2007yq}, which was also called the Flatt\'e model in Ref.~\cite{Artoisenet:2010va}. The renormalization of the two models has been discussed in Refs.~\cite{Braaten:2007nq,Artoisenet:2010va}, which shows that the cutoff dependence of the $T$-matrix can be eliminated. Therefore, we only take $\Lambda=500$ MeV in the following calculation. In principle, one can adopt the general models combining the model-I and model-II~\cite{Artoisenet:2010va}. In this work, in order to reduce the unknown parameters, we adopt the two models separately.

\begin{figure}[htp]
    \centering  \includegraphics[width=0.5\textwidth]{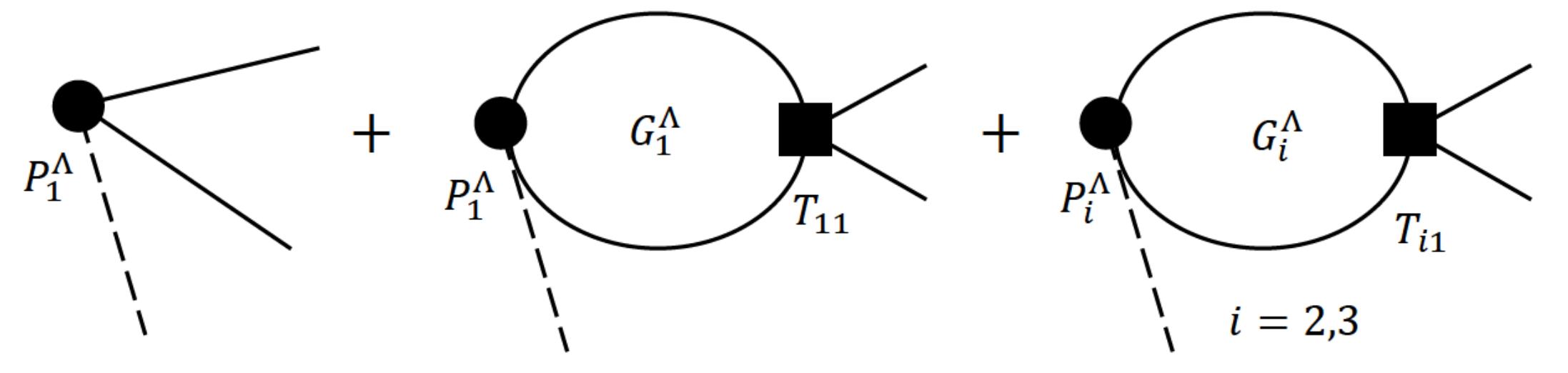}
    \caption{The Feynman diagrams for the $B^-\to J/\psi \Lambda \bar{p}$. The dashed lines represents the antiproton. The solid lines represent the three channels in Eq.~\eqref{eq:channels}. The $P_i^\Lambda$ are the vertices of the direct production. The solid square are the scattering $T$-matrix in Eq.~\eqref{eq:LSE}. The superscripts $\Lambda$ label the quantities with the cutoff-dependence.}\label{fig:feyn}
\end{figure}

We can calculate the amplitude of $B^-\to J/\psi \Lambda \bar{p}$ according to the Fig.~\ref{fig:feyn},
\begin{eqnarray}
    &&P_{1}^{\Lambda}+P_{1}^{\Lambda}G_{1}^{\Lambda}T_{11}+\sum_{i=2,3}P_{i}^{\Lambda}G_{i}^{\Lambda}T_{i1}\nonumber \label{eq:prod}\\
    &=&P_{1}^{\Lambda}(V_{11}^{\Lambda})^{-1}T_{11}+\sum_{i=2,3}\left(P_{i}^{\Lambda}-P_{1}^{\Lambda}(V_{11}^{\Lambda})^{-1}V_{1i}^{\Lambda}\right)G_{i}^{\Lambda}T_{i1}\nonumber \\
    &=&P_{1}T_{11}+\sum_{i=2,3}P_{i}T_{i1} ,
\end{eqnarray}
where we use the $T_{11}=V_{11}^{\Lambda}(1+G_{1}^{\Lambda}T_{11})+\sum_{i=2,3}V_{1i}^{\Lambda}G_{i}^{\Lambda}T_{i1}$ to obtain the first equation. The direct production vertices $P_i^\Lambda$ is cutoff-dependent. We can eliminate the cutoff-dependence by renormalizing the $P_i$  and get the final results following Ref.~\cite{Dong:2020hxe}.

\begin{figure*}[htp]
    \centering  \includegraphics[width=1.0\textwidth]{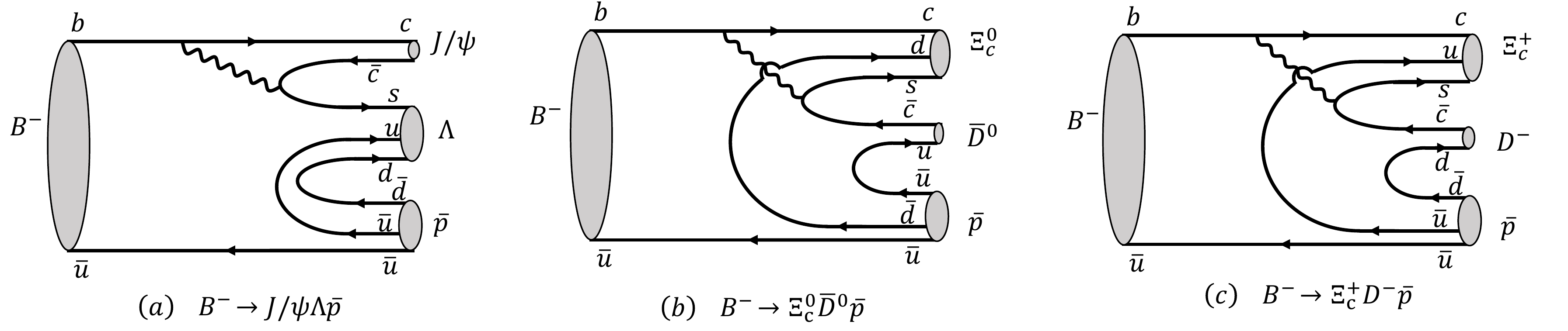}
    \caption{The Feynman diagrams for the $B^-\to J/\psi \Lambda \bar{p}$, $B^-\to J/\psi \Xi_c^0\bar{D}^0$ and $B^-\to J/\psi \Xi_c^+ D^-$. The wiggle lines represent the $W$ bosons. The three diagrams correspond to the $P_i^\Lambda$ vertices in Fig.~\ref{fig:feyn}. The rescattering effects of the hadrons are not included.}\label{fig:weakdecay}
\end{figure*}

The three direct production vertices $P_i^\Lambda$ from Eq.~\eqref{eq:prod} are presented at the quark level in Fig.~\ref{fig:weakdecay}. For the $P_2^\Lambda$ and $P_3^\Lambda$, their relative sign is very important.  In the isospin limit, there are relations $T_{21}=-T_{31}$ and $G^\Lambda_2=G^\Lambda_3$. Therefore, if there is the relation $P_2^\Lambda=P_3^\Lambda$, the contribution from the third diagram in Fig.~\ref{fig:feyn} will be canceled out for $i=2$ and $3$. In order to evaluate the ratio of $P_2^\Lambda/P_3^\Lambda$, we first define the initial state of the strong interaction,
\begin{equation}
    |\text{initial}\rangle=\left|\left\{[c(s\bar{c})_{1_{c}}^{1_{s}}]_{3_{c}}^{1/2_{s}}\bar{u}\right\}_{1_{c}}^{0_{s}}(q\bar{q})_{1_{c}}^{0_{s},0_{I}}(q\bar{q})_{1_{c}}^{0_{s},0_{I}}\right\rangle,
\end{equation}
where the superscripts represent the spin and isospin,  and the subscripts represent the color representation. The initial state is the spin-flavor-color wave functions of the $cs\bar{c}\bar{u}$ after weak vertices and two quark pairs generated from QCD vacuum. We estimate the $P_2^\Lambda/P_3^\Lambda$ by evaluating the overlap of the initial state and $|\Xi_c\bar{D}\bar{p}\rangle$,
\begin{equation}
    \frac{P_{2}^{\Lambda}}{P_{3}^{\Lambda}}=\frac{\langle\text{initial}|\Xi_{c}^{0}\bar{D}\bar{p}\rangle}{\langle\text{initial}|\Xi_{c}^{+}D^{-}\bar{p}\rangle}=-1.
\end{equation}
The above result indicates that there is no cancellation in the third diagrams of Fig.~\ref{fig:feyn}.

We can use $P_1 T_{11}+ P_2 (T_{21}-T_{31})$ to evaluate the lineshape of $J/\psi \Lambda$. However, the $P_1$  and $P_2$ are unknown parameters. In the following analysis, we will assume the $B^- \to J/\psi \Lambda \bar{p}$ is either the $J/\psi \Lambda$-driving or the $\Xi_c \bar{D}$-driving. Thus, we can only focus on $|T_{11}|^2$ and $|T_{21}-T_{31}|^2$ respectively to draw the lineshapes.

\section{Lineshapes}~\label{sec:lineshape}

In our calculation, the reflection principle in Eq.~\eqref{eq:reflection} is satisfied. The poles appear in pairs as Eq.~\eqref{eq:twopole}. There are three unknown parameters in each model, $c_0$, $c_1$ and $\tilde{c}$ for model-I and $m_0$, $g$ and $\tilde{c}$ for model-II. In order to fix the three parameters, we take a two-step procedure. First, we solve a two-channel problem ($|2\rangle $ and $|3\rangle$) to find poles at $M_\text{re}\pm\textrm{i}M_\text{im}\times 0.8$. We assume the main properties of the resonance (central mass and 80\% of the width) are determined by the $\Xi_c\bar{D}$ interaction as we explained in Sec.~\ref{sec:uniform}. In this step, we can determine the $c_0$ and $c_1$ for model-I, and $g$ and $m_0$ for model-II. In the second step, we will include the channel $|1\rangle$ and tune the $\tilde{c}$ to add the extra $20\%$ width. The partition of the contribution to the width can be slightly different but the $\Xi_c\bar{D}$ ones should be dominant. The two-step procedure can help us fix all the model parameters except the poles below the $\Xi_c^0\bar{D}^0$ threshold in sheet $(-,+,+)$, which correspond to the $\Xi_c \bar{D}$ bound states. For the $\Xi_c \bar{D}$ bound states, we take the vanishing widths in the first step and introduce  the width by coupling with the $J/\psi \Lambda$ channel in the second step. For the bound state solution, we have to eliminate one parameter manually in the first step. We take $c_1=0$ to ignore the effect of the isospin triplet channel in the model-I and take the BW mass as the bare mass $m_0=4338.3$ MeV in the model-II.

In determining the unknown parameters, we also use some criteria to delete the solutions that are inconsistent with the present understandings.
\begin{itemize}
    \item For model-I, we only keep the solutions with $c_0<0$ and $c_1>0$. Because the calculations in Refs.~\cite{Wang:2019nvm,Chen:2021cfl,Dong:2021juy} imply that the interactions for the isospin singlet and triplet of the $\Xi_c\bar{D}$ are attractive and repulsive, respectively.
    \item For model-II, we only keep the solution  with $m_0$ close to the present $\pcs4338$ mass, see $4238 \text{ MeV}<m_0<4438$ MeV.
    \item For the two-channel Flatt\'e model, there are in general four pole solutions belonging to two pairs considering the reflection principle as shown in Appendix~\ref{app:two_model}. We have checked that only one pair of them could appear in our region of interest. The other pair of solutions are about 2 GeV away from the $\pcs4338$ mass or even further.
    \item We neglect all the solution at sheet $(-,+,-)$. The numeral calculations show that the $T$-matrix with poles on this sheet is suppressed by two orders as compared to others.
\end{itemize}

In Table~\ref{tab:poleallowed}, we list the poles allowed by the above criteria. For the $(-,+,+)$ sheet, the poles are only allowed below the thresholds. For the $(-,-,+)$ sheet, all the poles are admitted except poles with $\Gamma=2$ MeV in the model-I. The poles on the sheet $(-,-,-)$ are only allowed by model-II.

\begin{table}
    \centering
    \caption{The allowed poles on different Riemann sheets for the pole masses in Eq.~\eqref{eq:polemass}. We use the ``o" to label the poles which are permitted by tuning the parameters in a reasonable range. The results for model-I and model-II are separated by ``/".}~\label{tab:poleallowed}
\begin{tabular}{ccccc}
    \hline \hline
    Model-I/II &
    $(-,+,+)$ &
    $(-,-,+)$ &
    $(-,-,-)$ &
    $(-,+,-)$\tabularnewline
    \hline
    $M_{0,1}$ &
    o/o &
    /o &
    /o &
    \tabularnewline
    $M_{0,2}$ &
    o/o &
    o/o &
    /o &
    \tabularnewline
    $M_{0,3}$ &
    o/o &
    o/o &
    /o &
    \tabularnewline
    \hline
    $M_{1,1}$ &
    &
    /o &
    /o &
    \tabularnewline
    $M_{1,2}$ &
    &
    o/o &
    /o &
    \tabularnewline
    $M_{1,3}$ &
    &
    o/o &
    /o &
    \tabularnewline
    \hline
    $M_{2,1}$ &
    &
    /o &
    /o &
    \tabularnewline
    $M_{2,2}$ &
    &
    o/o &
    /o &
    \tabularnewline
    $M_{2,3}$ &
    &
    o/o &
    /o &
    \tabularnewline
    \hline
    $M_{3,1}$ &
    &
    /o &
    /o &
    \tabularnewline
    $M_{3,2}$ &
    &
    o/o &
    /o &
    \tabularnewline
    $M_{3,3}$ &
    &
    o/o &
    /o &
    \tabularnewline
    \hline \hline
\end{tabular}
\end{table}

\begin{figure}[htp]
    \centering  \includegraphics[width=0.35\textwidth]{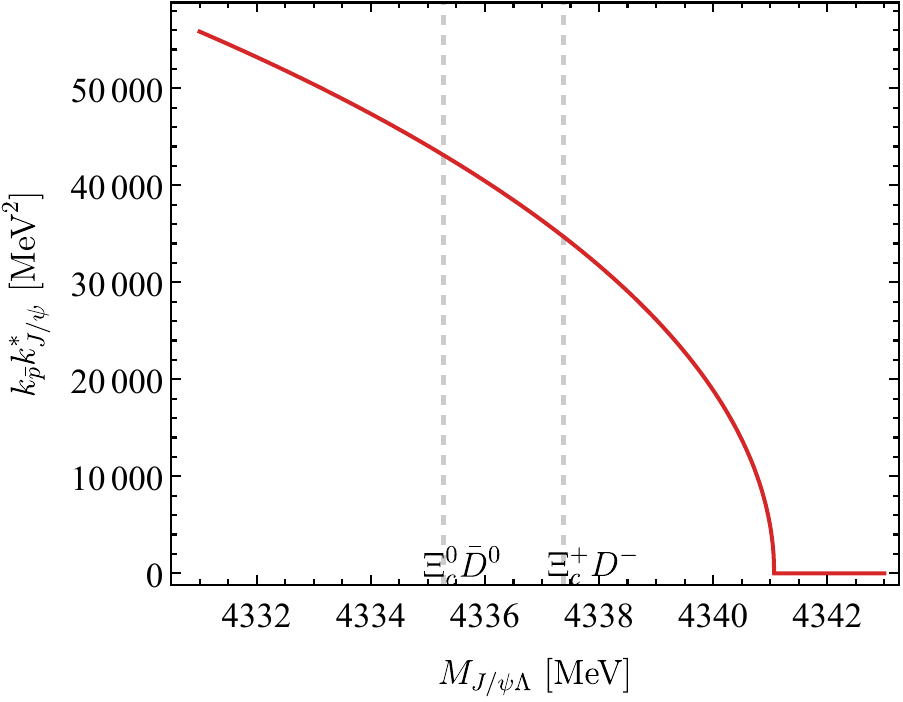}
    \caption{The lineshape of phase space of  $d\Gamma/ d M_{J/\psi \Lambda }$, which is proportional to $k_{\bar{p}} k^*_{J / \psi}$.  The dashed gray lines represent the $\Xi_c^0\bar{D}^0$ and $\Xi_c^+\bar{D}^-$ thresholds.}\label{fig:pphsp}
\end{figure}

We will focus on the lineshape of $J/\psi \Lambda$ in a narrow region near the $\Xi_c\bar{D}$ threshold in the decay $B^-\to J/\psi \Lambda \bar{p}$. Apart from the dynamical part, the phase space part reads
\begin{equation}
    {d\Gamma \over d M_{J/\psi \Lambda }} \propto k_{\bar{p}} k^*_{J / \psi},
\end{equation}
where $k_{\bar{p}}$ is the momentum of the $\bar{p}$ in the frame of the static $B^-$, and $k^*_{J/\psi}$ is the momentum of $J/\psi$ or $\Lambda$ in their center of mass frame. One can get their relations to $M_{J/\psi \Lambda} $ from Eq.~\eqref{eq:k}. The shape is shown in Fig.~\ref{fig:pphsp}. Because the mass of $B^-$ is very close to the three-body threshold of $ \Xi_c \bar{D}\bar{p}$, the kinetic-allowed phase space will fall dramatically in the region of interest. However, the descending behaviour has been included in experimental fitting. To investigate the possible bias of the experimental analysis, we will first show the lineshapes of the dynamical parts, i.e. $|T_{11}|^2$ and $|T_{12}-T_{13}|^2$ in the $J/\psi \Lambda$-driving and $\Xi_c\bar{D}$-driving mechanisms, respectively. The sole dynamical part could help show its differences with the BW parameterization. After that, we will take the phase space into consideration.

\subsection{Poles on the Sheet $(-,+,+)$}

\begin{figure*}[htp]
    \centering  \includegraphics[width=1.0\textwidth]{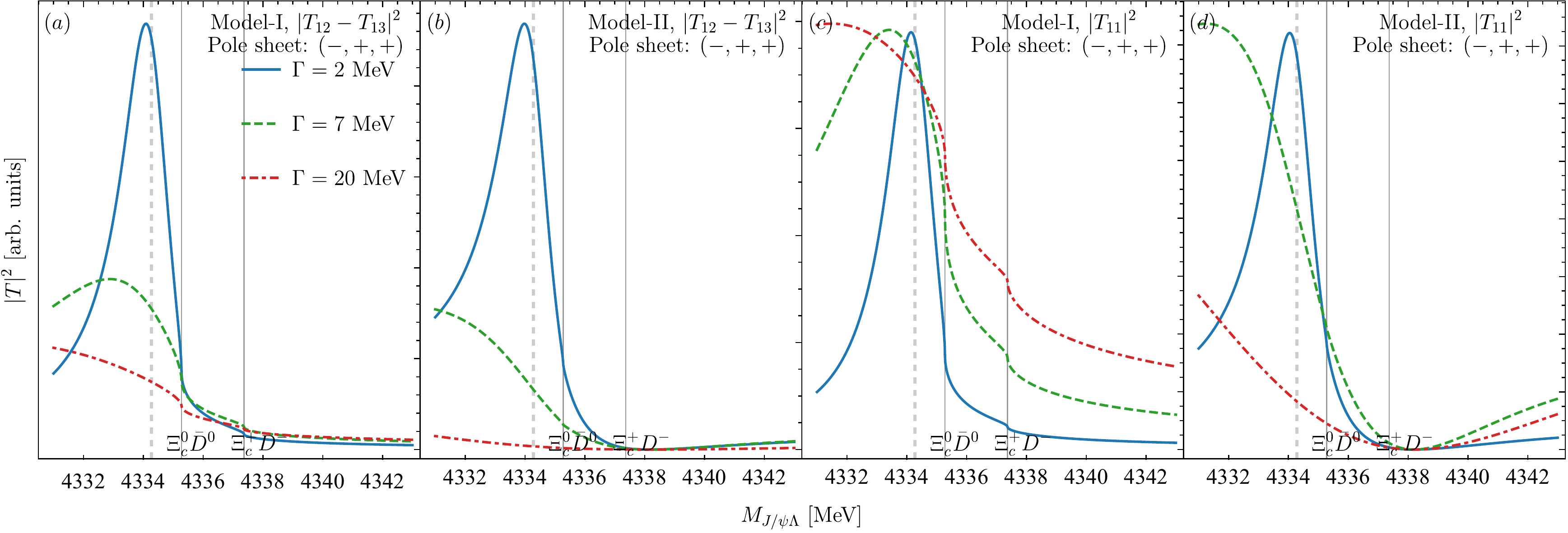}
    \caption{The $|T|^2$ with the poles on the $(-,+,+)$ sheet. The dashed gray line represents the central mass of the pole. The solid vertical lines represent two $\Xi_c\bar{D}$ thresholds. }\label{fig:popp}
\end{figure*}

In Fig.~\ref{fig:popp}, we plot the $|T|^2$ with the poles on the $(-,+,+)$ sheet.  The poles correspond to the bound states of $\Xi_c\bar{D}$. The widths come from the allowed decay $P_{\psi s}^\Lambda \to J/ \psi \Lambda$. One can see that for the poles with a small $\Gamma$ ($2$ MeV), the peaks almost appear at the central mass, which agrees with that in Fig.~\ref{fig:polecontri}. For the poles with a larger $\Gamma$, the peaks will deviate from the central mass.  One can see that the model-II gives the larger deviations than the model-I. Qualitatively, the two models with two production mechanisms still give the consistent lineshape. In all cases, the peaks appear below the lower threshold, which is different from the experimental pattern of the $\pcs4338$ peak that is close to the higher threshold, e.g. see Fig.~\ref{fig:exlineshape}. Therefore, the present experimental results seem to disfavor the $\pcs4338$ as the $\Xi_c \bar{D}$ bound state.

\subsection{Poles on the Sheet $(-,-,+)$}

\begin{figure*}[htp]
    \centering  \includegraphics[width=1.0\textwidth]{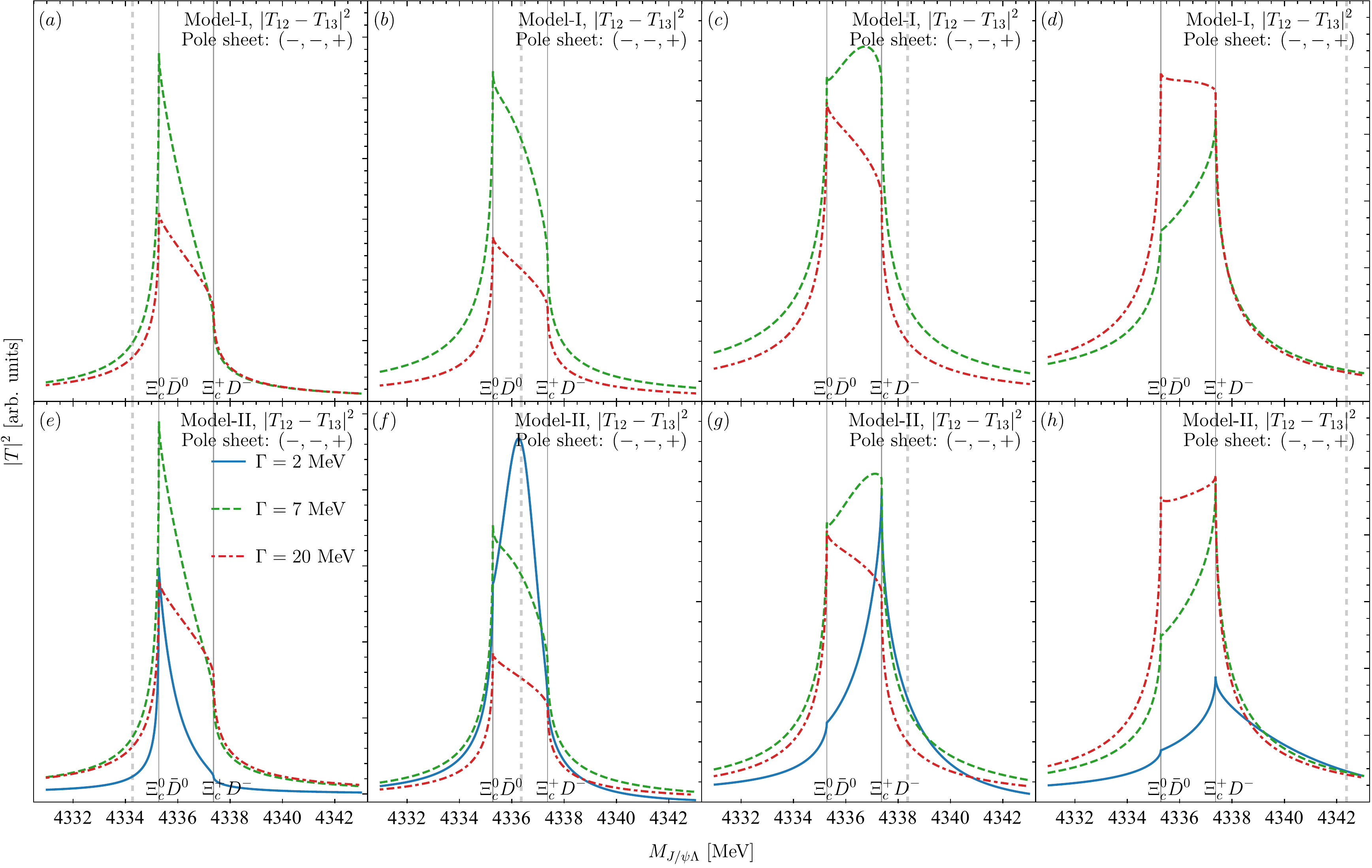}
    \caption{The $|T|^2$ with the poles on the $(-,-,+)$ sheet assuming the $\Xi_c\bar{D}$-driving production mechanism. The dashed gray line represents the central mass of the pole. The solid vertical lines represent two $\Xi_c\bar{D}$ thresholds.}\label{fig:pompa}
\end{figure*}

\begin{figure*}[htp]
    \centering  \includegraphics[width=1.0\textwidth]{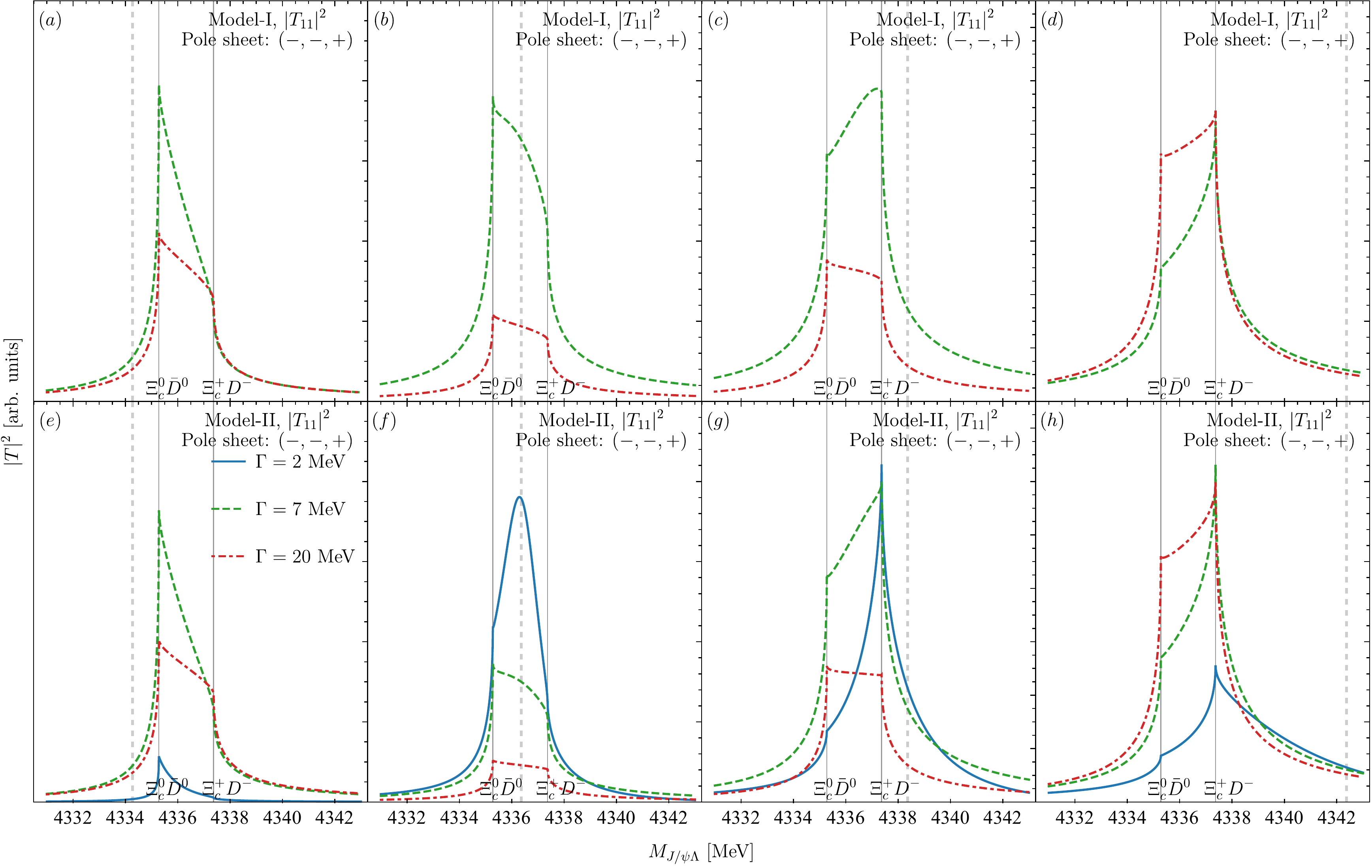}
    \caption{The $|T|^2$ with the poles on the $(-,-,+)$ sheet assuming the $J/\psi\Lambda$-driving production mechanism. The dashed gray line represents the central mass of the pole. The solid vertical lines represent two $\Xi_c\bar{D}$ thresholds.}\label{fig:pompb}
\end{figure*}

In Figs.~\ref{fig:pompa} and \ref{fig:pompb}, the $|T|^2$ with poles on the $(-,-,+)$ sheet through the $\Xi_c\bar{D}$-driving  and $J/\psi\Lambda$-driving mechanisms are presented, respectively. One can see that the lineshapes of resonances with the same pole mass in two models with different driving mechanisms are very similar. The only difference is that there do not exist solutions with $\Gamma=2$ MeV for model-I. In other words, it is hard to discern the dynamical model-I and model-II from the lineshapes in such a narrow energy range.

When the poles move from the positions below the lower threshold to the positions above the higher threshold, the peaks of $|T|^2$ will move from the lower threshold to the higher one as expected in Fig.~\ref{fig:polecontri}.  For the poles below the lower threshold (above the higher threshold), the peaks will not further go with the poles but appear at the the lower (higher) threshold. Most peaks appear as the  ``cusps" with unsmooth maximum points on the thresholds, which is obviously different from the smooth BW distribution.

In addition to the fact that the peak position does not reflect the $M_\text{re}$, the nominal half-widths of the enhancements could not correspond to the $M_\text{im}$. The shapes of the enhancement in the subfigures (c), (d), (g), (h) of Figs.~\ref{fig:pompa} and~\ref{fig:pompb} with $\Gamma=7$ and 20 MeV are constrained by the two thresholds in the energy extension rather than the $M_\text{im}$.  From Fig.~\ref{fig:region} and Fig.~\ref{fig:unif2} (b), one can see the closest physical regions of poles on the $(-,+)$ sheet are $(b),(c),(d)$, i.e. the two thresholds and intermediate regions.

\begin{figure}[htp]
    \centering  \includegraphics[width=0.45\textwidth]{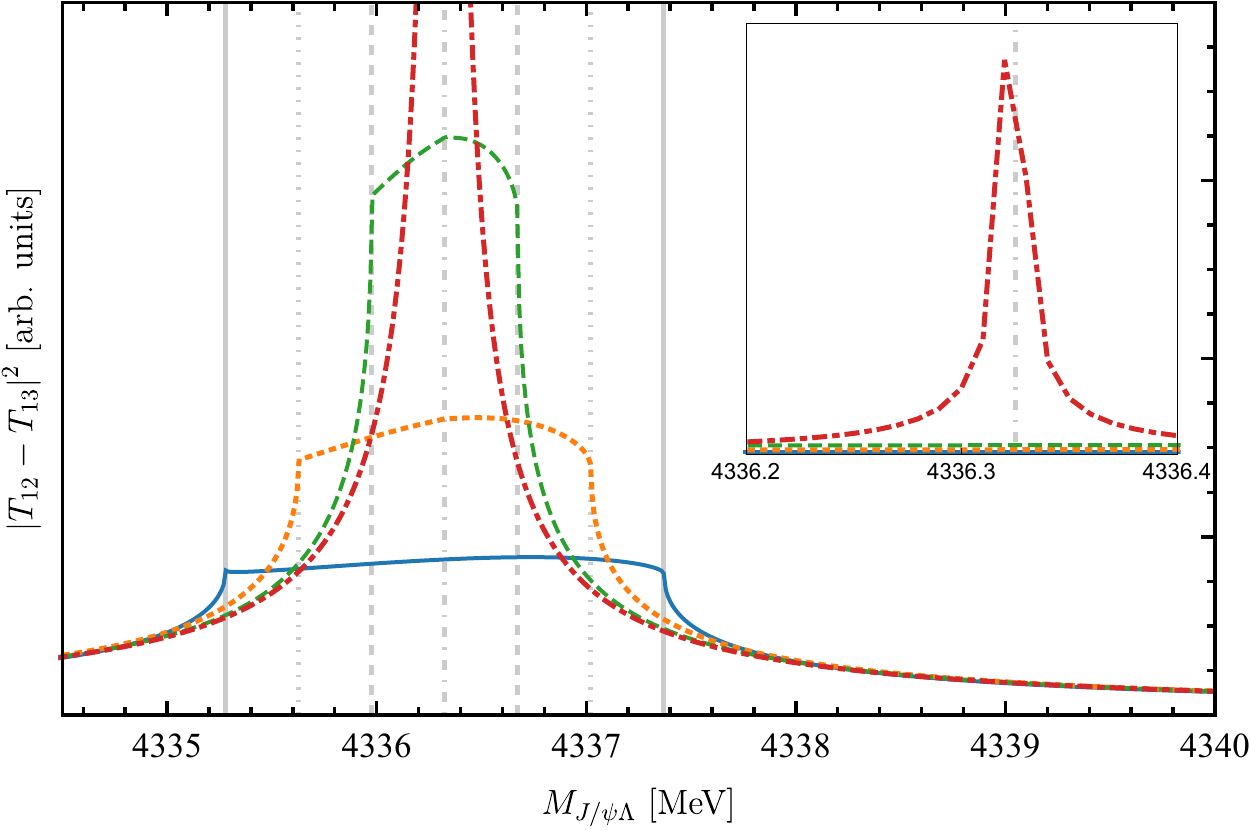}
    \caption{The $|T|^2$ changing with the $\Xi_c\bar{D}$ thresholds. The interaction permits a pole in $M_{2,2}$ on the $(-,-,+)$ sheet. The difference of the two thresholds is shrunk to zero by three equal steps. The blue lines is the zoomed version of the green dashed line in Fig.~\ref{fig:pompa} (c). The lineshapes and their corresponding thresholds are represented by the same types of lines. The red line is the one in the isospin symmetry limit, of which the full view is displayed in the small figure.  }\label{fig:change_th}
\end{figure}

 We use the interactions permitting a pole at $M_{2,2}$ and giving  the $|T|^2$ with $\Gamma=7$ MeV in Fig.~\ref{fig:pompa} as an example to investigate the effect of the double thresholds. We keep the same interaction and shrink the differences of two thresholds by three equal steps. We compare the $|T|^2$ with different thresholds in Fig.~\ref{fig:change_th}. One can see that the energy extension will be narrowed with the change of the thresholds. Finally, the enhancement becomes a very sharp peak when the two thresholds become the same in the isospin symmetry limit. For the sharp peak, the channels with different isospins are decoupled. We find the peak corresponds to a near-threshold virtual state of $|\Xi_c\bar{D},I=0\rangle$.  Therefore, the shapes in the subfigures (c), (d), (g), (h) of Figs.~\ref{fig:pompa} and~\ref{fig:pompb} are the virtual states broadened by the splitting of the thresholds. The states coupling to two thresholds lead to a different pattern from those with a single threshold effect.

The peaks of the lineshapes could truly reflect the pole position only when the poles on the $(-,-,+)$ sheet lie between the two thresholds, and its $\Gamma$ is small, such as the lines coinciding to $\Gamma=2$  MeV in Fig.~\ref{fig:pompa} (f) and Fig.~\ref{fig:pompb} (f). Otherwise, the BW parameterization could introduce a large discrepancy. In principle, the experimental $\pcs4338$ could arise from a pole higher than the peak position with the lineshape distorted by the double threshold effect.

\subsection{Poles on the Sheet $(-,-,-)$}

\begin{figure*}[htp]
    \centering  \includegraphics[width=1.0\textwidth]{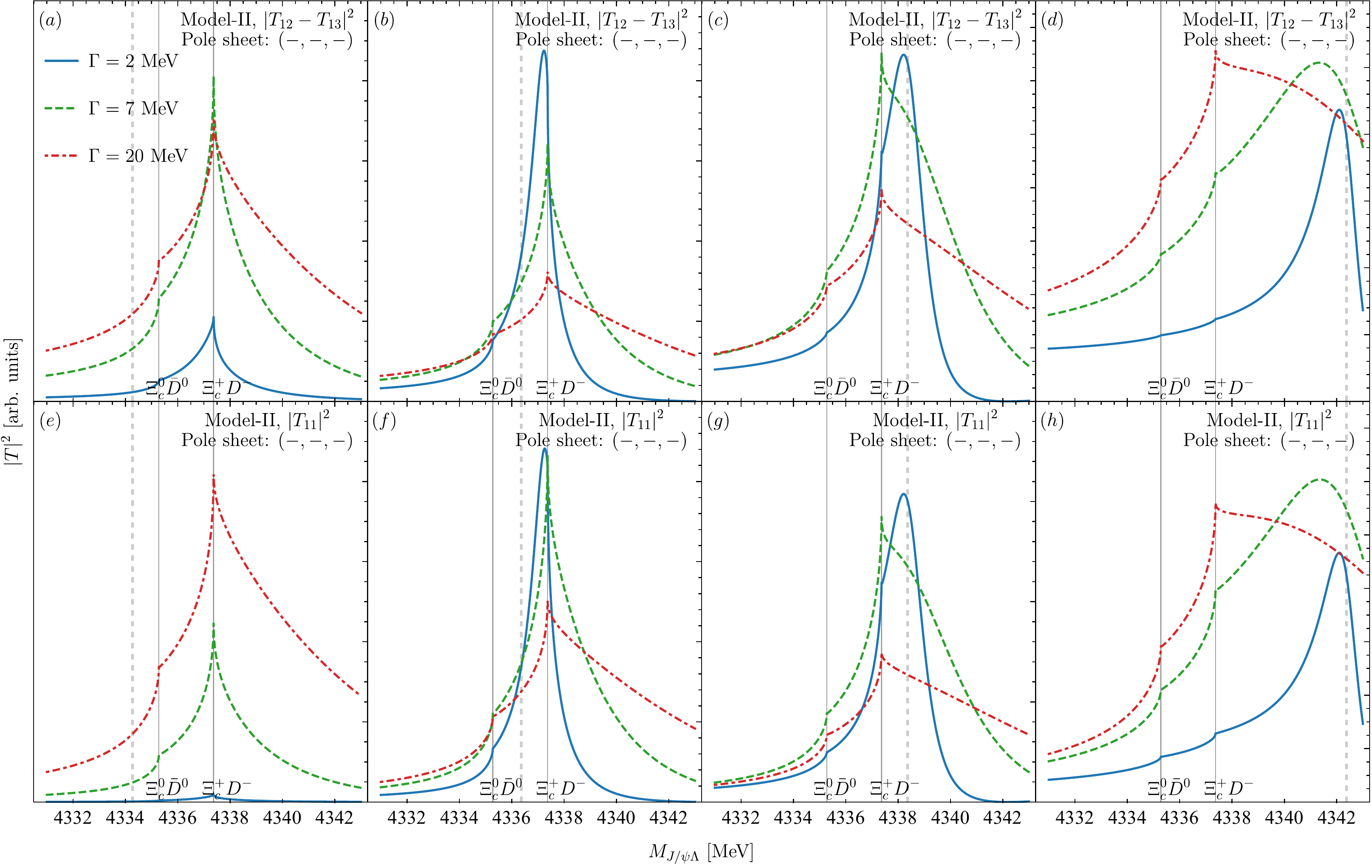}
    \caption{The$|T|^2$ with the poles on the $(-,-,-)$ sheet. The dashed gray line represents the central mass of the pole. The solid vertical lines represent two $\Xi_c\bar{D}$ thresholds.}\label{fig:p0mm}
\end{figure*}

In Fig.~\ref{fig:p0mm}, we show the $|T|^2$ with poles on the sheet $(-,-,-)$. One can see for the poles below the higher threshold, the peaks tend to appear at the higher thresholds. For the poles above the higher threshold, peaks tend to move with the pole position. The picture is in good agreement with that in Fig.~\ref{fig:polecontri}. In other words, the experimental enhancement of $\pcs4338$ could arise from a pole on the sheet $(-,-,-)$ below the higher threshold even below the first threshold.  Considering the resolution of the detector, the ``cusp" at the higher threshold  amplified by the pole in other places would be identified as a BW peak and give a misleading resonance mass and lifetime. Only when the poles on the sheet $(-,-,-)$ are above the higher threshold and with the small $\Gamma/2$ comparing with the $|M_\text{re}-M_{T_2}|$, the effects from the thresholds could be less important. In this way, the lineshapes could suit to the BW parameterization, such as ones in Fig.~\ref{fig:p0mm} (c) and (g) with $\Gamma=2$ MeV, and the ones in Fig.~\ref{fig:p0mm} (d) and (f) with $\Gamma=2$ and $7$ MeV.

\subsection{Lineshapes considering the phase space }
\begin{figure*}[htp]
    \centering  \includegraphics[width=1.0\textwidth]{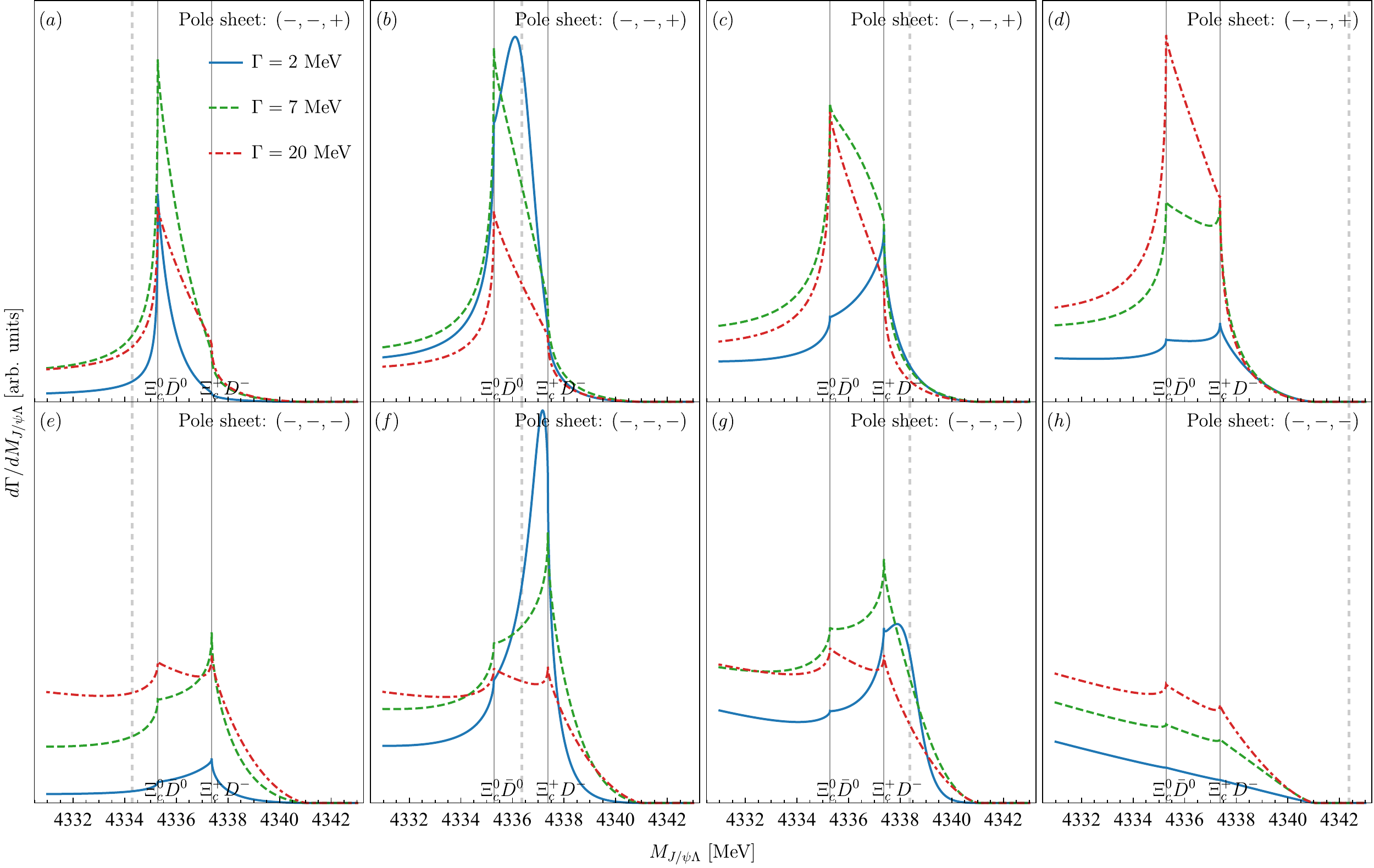}
    \caption{The $J/\psi \Lambda $ invariant mass spectrum in the decay $B^-\to J/\psi \Lambda \bar{p}$ with the poles on the $(-,-,-)$ sheet and $(-,-,+)$ sheet. The dashed gray line represent the central mass of the pole. The solid vertical lines represent two $\Xi_c\bar{D}$ thresholds.}\label{fig:plineshape}
\end{figure*}

In this part, we take the phase space in Fig.~\ref{fig:pphsp} into consideration. Since the two models and two driving mechanisms give the similar shapes in the dynamical part, we only show the lineshape using the model-II in the $\Xi_c\bar{D}$-driving mechanism as an example. In present models, the poles on the $(-,+,+)$ sheet can not give a peak above the second threshold. Therefore, we only draw the lineshapes corresponding to the poles on the $(-,-,+)$ ans $(-,+,+)$ sheets as shown in Fig.~\ref{fig:plineshape}. Apparently, the descending phase space will distort the lineshapes. We still see the enhancements or cusps in the vicinity of the $\Xi_c\bar{D}$ thresholds.

\section{Summary}~\label{sec:sum}

In this work, we explore the possible effect of the $\Xi_{c}^+D^-$ and $\Xi_c^0 \bar{D}^0$ thresholds on the lineshapes of the $\pcs4338$ in the $J/\psi \Lambda$ invariant mass spectrum of the $B^-\to J/\psi \Lambda \bar{p} $. We assume the properties of $\pcs4338$ are mainly determined by the $\Xi_c\bar{D}$ effect with the coupling to $J/\psi \Lambda$ as a small correction.  With the knowledge of the topological structure of the two-channel system ($\Xi_{c}^+D^-$ and $\Xi_c^0 \bar{D}^0$ channels), we use a qualitative picture of the uniformization to understand the positions of the peaks. We use two dynamical models to calculate the  $J/\psi \Lambda$ lineshapes in the LSE formalism to verify the picture. In the calculation, the unitarity and the analyticity associated with the two-channel system are considered carefully. The formalism is equivalent to the $K$-matrix parameterization in RPP~\cite{ParticleDataGroup:2020ssz}. The model-I is the zero-range model and the model-II is equivalent to the Flatt\'e model. We consider two different production mechanisms in $B^-\to  J/\psi \Lambda \bar{p}$, $J/\psi \Lambda$-driving one and $\Xi_c \bar{D}$ one. We adopt the twelve pole masses on $(-,+,+)$, $(-,-,+)$ and $(-,-,-)$ sheets to investigate the specific lineshapes. As a by-product, we calculate the ratio of the isospin violation decays of $\pcs4338$.

From the numerical results, we can obtain the following conclusions:
\begin{itemize}
\item The two models and two production mechanisms permit the similar lineshapes. The lineshapes near the $\Xi_c\bar{D}$ thresholds cannot be used to discern the underling dynamics. Because the lineshapes are roughly fixed by the analyticity and unitarity, in particular the positions of the poles and branch cuts.
\item For the poles near the thresholds, the conventional BW parameterization cannot uncover the real pole mass. Instead, the parameterization in Eq.~\eqref{eq:twopole} with variable $z$ after uniformization  gives consistent lineshapes with the two dynamical models.  The peaks of poles tend to appear in the closest point in physical region of the pole in the uniformized $z$-plane as shown in Fig.~\ref{fig:region} rather than the central mass position predicted by the BW parameterization.
\item For the poles on the $(-,-,+)$ sheet, one could observe that the enhancements with widths are determined by the threshold differences rather than the imaginary parts of the pole masses. The lineshapes are the virtual state peaks in the isospin limits which are broadened  by the isospin splitting of the two thresholds. The phenomena are consequences of the double thresholds appearing near the resonance poles.
\item The enhancement of $\pcs4338$ could potentially arise from the pole on the $(-,-,+)$ sheet  well above the $\Xi_c^+D^-$ threshold and the pole on the $(-,-,-)$ plane well below the $\Xi_c^+D^-$ threshold, where the BW parameterization may be misleading.
\item As a by-product, the isospin violating decay ratio ${\Gamma_{P_{\psi s}^{\Lambda}\to J/\psi\Sigma}}/{\Gamma_{P_{\psi s}^{\Lambda}\to J/\psi\Lambda}}$ could be at most $10\%$.
\end{itemize}
Considering the potential discrepancy of BW parameterization, we urge the LHCb Collaboration to perform the analysis with the $K$-matrix parameterization, or the ``Flatt\'e" parameterization when the  amount of data becomes large enough. The present analysis using uniformization is very general, which can be used to understand the other near-threshold structures.

\begin{appendix}

\section{Zero-range model and Flatt\'e model}~\label{app:two_model}

In order to show the relations of model-I and model-II with the other similar ones in literature, we give the reduced non-relativistic $\Xi_c\bar{D}$ two-channel scattering formalism. Ignoring the effect of $J/\psi \Lambda$, we can solve the LSEs $\tilde{T}=\tilde{V}+\tilde{V}\tilde{G}\tilde{T}$, with
\begin{equation}
    \tilde{V}=\left[\begin{array}{cc}
        V_{22} & V_{23}\\
        V_{32} & V_{33}
    \end{array}\right],\quad\tilde{G}=\text{diag}\{G_{2},G_{3}\}.\label{eq:twoc_LSE}
\end{equation}
With the non-relativistic approximation, the $G_i$ reads
\begin{eqnarray}
    G_{i}(E)&\approx&\frac{1}{2m_{i1}m_{i2}}\int_{0}^{\Lambda}\frac{l^{2}dl}{(2\pi)^{2}}\frac{1}{E-m_{i1}-m_{i2}-{\bm{l}^{2} \over 2\mu_{i}}+i\epsilon}\nonumber\\ &\approx& N_{i}[\frac{2}{\pi}\Lambda+ik_{i}],
\end{eqnarray}
where the constant $N_i$ is  $N_{i}=-\frac{1}{2m_{i1}m_{i2}}\frac{\mu_{i}}{4\pi}\equiv N$. We have neglected the mass difference from the isospin violation in $N_i$. $k_i$ is defined as $k_{i}=\sqrt{2\mu_{i}(E-m_{i1}-m_{i2})+i\epsilon}$. The mass splittings of the two channels in the expression of $k_i$ are kept. Solving the LSEs, we can obtain the inverse of the amplitude $\tilde{\mathcal{A}}^{-1}$,
\begin{equation}
    \tilde{{\cal A}}^{-1}=N^{-1}\tilde{T}^{-1}=N^{-1}(\tilde{V}^{-1}-\tilde{G})=\tilde{K}^{-1}-ik_{i}\delta_{ij},~\label{eq:k-matrix}
\end{equation}
where $\tilde{K}^{-1}=N^{-1}\tilde{V}^{-1}-\frac{2}{\pi}\delta_{ij}\Lambda$. The Eq.~\eqref{eq:k-matrix} is the $K$-matrix parameterization in RPP~\cite{ParticleDataGroup:2020ssz}.

For model-I, one can define the cutoff-independent $\tilde{K}^{-1}$
\begin{eqnarray}
    \tilde{K}_{\text{I}}^{-1}=\left[\begin{array}{cc}
        \frac{1}{a_{22}} & \frac{1}{a_{23}}\\
        \frac{1}{a_{23}} & \frac{1}{a_{22}}
    \end{array}\right].
\end{eqnarray}
For one channel case, the $a_{ii}$ is the scattering length. Therefore, the above model is just the zero-range model~\cite{Braaten:2007nq,Artoisenet:2010va}.

For model-II, one can get the amplitude,
\begin{eqnarray}
\tilde{{\cal A}}_{22}&=&-\tilde{{\cal A}}_{23}=\tilde{{\cal A}}_{33}\\&=&-\frac{g^{2}/2}{4(s-m_{0}^{2})(m_{1}+m_{2})\pi+g^{2}\frac{2}{\pi}\Lambda+i(\frac{g^{2}k_{2}}{2}+\frac{g^{2}k_{3}}{2})}.\nonumber
\end{eqnarray}
The above expression is the Flatt\'e formalism~\cite{Flatte:1976xu}.
If one takes the approximation, $E+m_{0}\approx m_{1}+m_{2}$ and the following replacement,
\begin{equation}
\frac{g^{2}}{2}\to\frac{g'^{2}}{2}8(m_{1}+m_{2})^{2},
\end{equation}
one can get
\begin{eqnarray}
\tilde{{\cal A}}_{22}&=&-\tilde{{\cal A}}_{23}=\tilde{{\cal A}}_{33}\\&\approx&\frac{g'^{2}/2}{E-m_{0}+g'^{2}\frac{2}{\pi}\Lambda+i(\frac{g'^{2}k_{2}}{2}+\frac{g'^{2}k_{3}}{2})},
\end{eqnarray}
which is the Flatt\'e-like formalism used in Ref.~\cite{Hanhart:2007yq}.

If one replaces the $E$, $k_2$, $k_3$ in the  Flatt\'e model with $z$ in uniformization scheme, one can obtain the denominator,
\begin{equation}
    \text{Denominator}=A(z+{1\over z})^2+iB(z+{1\over z})+iC(z-{1\over z}) +D,
\end{equation}
 where $A,B,C,D$ are real parameters. Therefore, there are four pole solutions in the two-channel Flatt\'e model. Meanwhile, the four solutions satisfy the reflection rules in Eq.~\eqref{eq:reflection}, and can be classified into two groups.

\section{Isospin violating decay $\pcs4338 \to J/\psi \Sigma$}~\label{app:iso_v}

The peak of the $\pcs4338$ resonance is very close to the $\Xi_c^+ D^-$ threshold, which could make the $\pcs4338$ prefer to couple with the $\Xi_c^+ D^-$ rather than equally couple to two thresholds, $\Xi_c^+ D^-$ and $\Xi_c^0 \bar{D}^0$. A natural consequence of the above picture is the potentially large isospin violating decay $\pcs4338 \to J/\psi \Sigma$, as in the di-pion decays of the $X(3872)$~\cite{Li:2012cs,Wu:2021udi,Meng:2021kmi}. The isospin violating decays of the $X(3872)$ and $P_{\psi}^N(4457)$ were also investigated in Refs.~\cite{Gamermann:2009fv,Guo:2019fdo}. We can evaluate the isospin violating decay ratio within a two-channel coupling scenario.

\begin{figure}[htp]
    \centering  \includegraphics[width=0.4\textwidth]{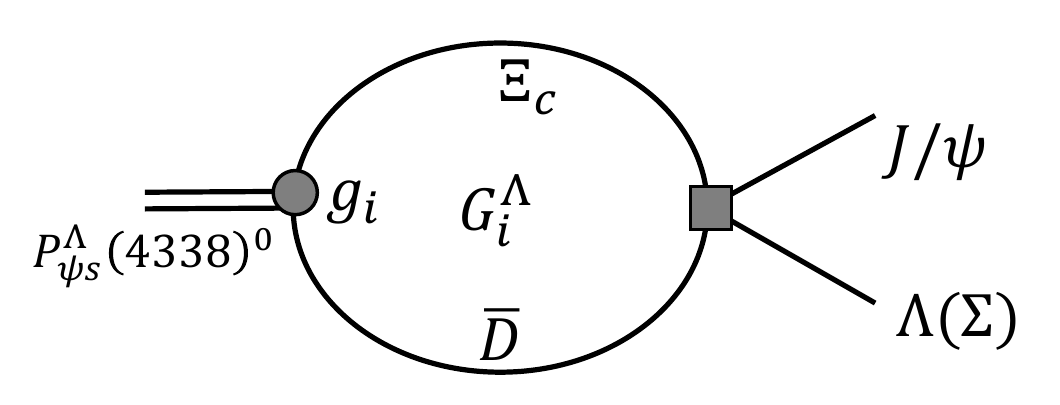}
    \caption{The decay diagram for $\pcs4338 \to J/\psi \Lambda (\Sigma) $.}\label{fig:iso_v_decay}
\end{figure}

We solve the LSEs in Eq.~\eqref{eq:twoc_LSE} in the first step of our two-step procedure. One can extract the couplings $g_2$ and $g_3$ from the residues of the $\tilde{T}$-matrix
\begin{equation}
    \lim_{E\to M_{\rm{pole}}}(E-M_{\rm{pole}})\tilde{T}_{ij}\propto g_{i}g_{j},
\end{equation}
where $i,j$ are the indices of the channels in Eq.~\eqref{eq:channels}. The Feynman diagram for the $\pcs4338\to J/\psi\Lambda (\Sigma)$ decay is presented in Fig.~\ref{fig:iso_v_decay}. We assume the vertices $\Xi_c\bar{D}-J/\psi\Lambda (\Sigma)$ satisfy the isospin symmetry. The isospin symmetry is violated in the $g_i$ couplings and $G_i^\Lambda$ due to the mass splitting and the phase space differences. Therefore, the isospin violating decay ratio reads
\begin{eqnarray}
R&=&\frac{\Gamma_{P_{\psi s}^{\Lambda}\to J/\psi\Sigma}}{\Gamma_{P_{\psi s}^{\Lambda}\to J/\psi\Lambda}}\\&=&\left|\frac{g_{3}G_{3}(M_\text{re})+g_{2}G_{2}(M_\text{re})}{g_{3}G_{3}(M_\text{re})-g_{2}G_{2}(M_\text{re})}\right|^{2}\times\frac{k(M_\text{re},m_{J/\psi},m_{\Sigma})}{k(M_\text{re},m_{J/\psi},m_{\Lambda})}.\nonumber
\end{eqnarray}

In Table~\ref{tab:isospin_vio}, we present the isospin violating decay ratios for the different pole masses in two models. For all the model-II results and model-I results with poles on the $(+,+)$ sheet, we set the interaction in the isospin triplet channel to be vanishing and only keep the isospin singlet interaction. This strategy will lead to the difference of $|g_2|$ and $|g_3|$ negligible. The isospin violating decays mainly stem from  $G_i^\Lambda$ due to the mass splittings and the phase space differences. Thus, one can see the ratios only depend on the central mass and are insensitive to the $\Gamma$ of the pole mass. Without the contribution of the difference of $|g_2|$ and $|g_3|$, the ratios are smaller by almost one order than others.

In the calculation, the largest isospin violating ratio is at the order of $10\%$. The ratio is small as compared to that of $X(3872)$, since the $\pcs4338$ is not as close to the threshold as the $X(3872)$ and is not as narrow as $X(3872)$. However, the $10\%$ isospin violating effect could still be detected in experiments.
\begin{table}
\centering
\caption{The isospin violating decay ratios for the different pole masses in the two models.}~\label{tab:isospin_vio}
\begin{tabular}{cccccc}
\hline
\hline
$R$ &
I: $(+,+)$ &
I: $(-,+)$ &
II: $(+,+)$ &
II: $(-,+)$ &
II$(-,-)$\tabularnewline
\hline
$M_{0,1}$ &
0.002 &
 &
0.002 &
0.002 &
0.002\tabularnewline
$M_{0,2}$ &
0.002 &
0.014 &
0.002 &
0.002 &
0.002\tabularnewline
$M_{0,3}$ &
0.002 &
0.019 &
0.002 &
0.002 &
0.002\tabularnewline
\hline
$M_{1,1}$ &
 &
 &
 &
0.007 &
0.007\tabularnewline
$M_{1,2}$ &
 &
0.043 &
 &
0.007 &
0.007\tabularnewline
$M_{1,3}$ &
 &
0.038 &
 &
0.007 &
0.007\tabularnewline
\hline
$M_{2,1}$ &
 &
 &
 &
0.002 &
0.002\tabularnewline
$M_{2,2}$ &
 &
0.069 &
 &
0.002 &
0.002\tabularnewline
$M_{2,3}$ &
 &
0.048 &
 &
0.002 &
0.002\tabularnewline
\hline
$M_{3,1}$ &
 &
 &
 &
0.001 &
0.001\tabularnewline
$M_{3,2}$ &
 &
0.092 &
 &
0.001 &
0.001\tabularnewline
$M_{3,3}$ &
 &
0.052 &
 &
0.001 &
0.001\tabularnewline
\hline
\hline
\end{tabular}
\end{table}

Apart from the above formalism, one can choose to use a four-channel coupling method to estimate the isospin violating effect. The interactions can be introduced as
\begin{eqnarray}
V_{\text{I}}=\frac{1}{2}\left[\begin{array}{cccc}
0 & 0 & -\tilde{c} & \tilde{c}\\
0 & 0 & \tilde{c} & \tilde{c}\\
-\tilde{c} & \tilde{c} & c_{1}+c_{0} & c_{1}-c_{0}\\
\tilde{c} & \tilde{c} & c_{1}-c_{0} & c_{1}+c_{0}
\end{array}\right],
\end{eqnarray}
or alternatively,
\begin{eqnarray}
V_{\text{II}}=\frac{1}{2}\left[\begin{array}{cccc}
0 & 0 & -\tilde{c} & \tilde{c}\\
0 & 0 & \tilde{c} & \tilde{c}\\
-\tilde{c} & \tilde{c} & \frac{g^{2}}{E^2-m_{0}^{2}} & -\frac{g^{2}}{E^2-m_{0}^{2}}\\
\tilde{c} & \tilde{c} & -\frac{g^{2}}{E^2-m_{0}^{2}} & \frac{g^{2}}{E^2-m_{0}^{2}}
\end{array}\right],
\end{eqnarray}
where we insert the $J/\psi \Sigma$ channel based on the Eqs.~\eqref{eq:model-I} and ~\eqref{eq:model-II}. One can use the two-step procedure to determine the three unknown parameters in each model. The couplings of $P_{\psi s}^\Lambda-J/\psi \Lambda $ and $P_{\psi s}^\Lambda-J/\psi \Sigma $ can be extracted from the residues of the four-channel $T$-matrix. The final results have no qualitative difference with those in Table~\ref{tab:isospin_vio}.
\end{appendix}


\begin{acknowledgements}
We are grateful to the helpful discussions with Xin-Zhen Weng and Guang-Juan Wang. L.M. is grateful to the Wren A. Yamada for the helpful discussions on uniformization. This project was supported by the National
Natural Science Foundation of China (11975033 and 12070131001). This
project was also funded by the Deutsche Forschungsgemeinschaft (DFG,
German Research Foundation, Project ID 196253076-TRR 110). B.W. was supported by
the National Natural Science Foundation
of China under Grant No. 12105072, the Youth Funds of Hebei Province (No. A2021201027), and the Start-up Funds for Young Talents of Hebei University
(No. 521100221021).
\end{acknowledgements}

\bibliography{ref}

\end{document}